\documentclass[12pt]{iopart}

\usepackage{iopams}  
\usepackage{graphicx}
\usepackage{hyperref,epstopdf}
\usepackage{xcolor}
\begin{document}

\title[]{Collective states of even-even nuclei in $\gamma$-rigid quadrupole Hamiltonian with Minimal Length  under  the sextic potential}

\author{ A. El Batoul , M. Oulne and I.Tagdamte}

\address{High Energy Physics and Astrophysics Laboratory, Department of Physics, Faculty of Sciences Semlalia, Cadi Ayyad University P.O.B 2390, Marrakesh 40000, Morocco}

\ead{elbatoul.abdelwahed@edu.uca.ma, oulne@uca.ac.ma, tagdamte.imad@gmail.com}
\vspace{10pt}

\begin{indented}
\item[]May 2021
\end{indented}

\begin{abstract}
	
In the present paper, we study the collective states of even-even nuclei in $\gamma$-rigid mode within the sextic potential and the Minimal Length (ML) formalism in Bohr–Mottelson model. The eigenvalues problem for this latter is solved by means conjointly of  Quasi-Exact Solvability (QES) and a Quantum Perturbation Method (QPM). Numerical calculations are performed for 35 nuclei: $  ^{98-108}$Ru,  $^{100-102}$Mo, $^{116-130}$Xe, $^{180-196}$Pt, $^{172}$Os, $^{146-150}$Nd, $^{132-134}$Ce, $^{154}$Gd, $^{156}$Dy and $^{150-152}$Sm. Through this study, it appears that our elaborated model leads to an improved agreement of the theoretical results with the corresponding experimental data by reducing the rms with a rate going up to 63\% for some nuclei. This comes out from the fact that we have combined the sextic potential, which is a very useful phenomenological potential, with the formalism of the ML which is based on the generalized uncertainty principle and which is in turn a quantum concept widely used in quantum physics. Besides, we investigate the effect of ML on energy ratios, transition rates, moments of inertia and a shape phase transition for the most numerous isotopic chains, namely Ru, Xe, Nd and Pt.
\end{abstract}

\section{Introduction}
It is well known that the collective motions are the most striking phenomenon, appearing widely in atomic, molecular and nuclear structure physics, which are traditionally described by means of collective models. One of the well-known example is the Bohr–Mottelson model (BMM)\cite{b1,b2} reflecting some theoretical aspects in the physical observables wherein a good agreement with experimental data has been observed\cite{b3}. In the same context, a theoretical group description of collective nuclear properties is alternatively afforded by the interacting boson model (IBM)\cite{b4,b5,b6,b7}, which is intrinsically a pair coupling model, with coherent pairs of monopolar and quadrupole fermions (nucleons: protons or neutrons) considered approximately as bosons. However, by using of an intrinsic coherent state approach\cite{b8,b9}, the dynamical IBM Hamiltonian can be expressed in terms of the geometrical shape variables of BMM. Thereby, the resulted collective differential equation of state is widely more involved, bearing only a marginal equivalence to the Bohr Hamiltonian\cite{b10,b11,b12}. In reality, this is not surprising because of the conceptual distinction that exists between the two models: the BMM is considered purely geometrical, while IBM is intuitively a large truncation of the shell model, where the bound states with low energy and positive parity are defined by the interaction of bosons $s$ and $d$. On the one hand, the connection with the shell model is made by supposing that these bosons $s$ and $d$ correspond to correlated pairs of valence nucleons coupled to $J^+=0^+$ and $J^+=2^+$.  Nonetheless, in the previous efforts to  connect the two approaches\cite{b4,b9,b10}, it turned out that the limiting dynamical symmetries U(5)\cite{b5}, SU(3)\cite{b13}, and O(6)\cite{b6}, identified as subgroup chains of the SU(6) symmetry of IBM, have similarities in BMM pictured by its solvable instances  corresponding to  the nuclear shape phases describing a spherical vibrator\cite{b1}, an axially symmetric rotor\cite{b2}, and a $\gamma$-soft rotor\cite{b14} respectively. However, alternative solutions were proposed in parallel with the occurrence of critical point symmetries. As examples, we can cite : E(5)\cite{b15}, X(5)\cite{b16}, Y(5)\cite{b17} and Z(5)\cite{b18}, which describe the nuclei situated in the critical points of the shape phase transitions from spherical vibrator [U(5)] to a  $\gamma$-unstable [O(6)] nuclei, from spherical vibrator to prolate rotor [SU(3)], from axial rotor to triaxial rotor, and from prolate rotor to oblate rotor, respectively. These CPSs are constructed from the Hamiltonian by separating the collective variables ($\beta$ and $\gamma$) defining the deviation from sphericity and axiality, respectively. 
\par We recall that the quadrupole Hamiltonian of Bohr in its original form is written as\cite{b1,b2} :

\begin{equation}
\fl H_{B}=-\frac{\hbar^{2}}{2 B}\bigg[\frac{1}{\beta^{4}} \frac{\partial}{\partial \beta} \beta^{4} \frac{\partial}{\partial \beta}+\frac{1}{\beta^{2} \sin 3 \gamma} \frac{\partial}{\partial \gamma} \sin 3 \gamma \frac{\partial}{\partial \gamma}-\frac{1}{4 \beta^{2}} \sum_{k=1,2,3} \frac{Q_{k}^{2}}{\sin ^{2}\left(\gamma-\frac{2}{3} \pi k\right)}\bigg]+V(\beta, \gamma) ,
\end{equation}
where $ Q_{k} $ ($k=1,2,3$) are the components of angular momentum in the intrinsic frame and $ B $ is the mass parameter, which is usually considered constant. It should be noted that the number of degrees of freedom for this Hamiltonian is 5, but by imposing some constraints on its kinetic energy (for example by fixing the $\gamma$ variable to be equal to zero), one can reduce it to 3. The obtained result in this case corresponds to the $\gamma$-rigid mode\cite{b19} for nuclei. This is the mode on which our study is essentially based. So, in the present work we focused on a special class of solutions for the Bohr Hamiltonian in the presence of a Minimal Length (ML)\cite{b20} in X(3) a quasi-exactly solvable potential namely: Sextic potential\cite{b21}, where we have obtained the expressions of eigenvalues and wave functions by means of Quasi-exact solvability (QES)\cite{b21} and a quantum perturbation method(QPM)\cite{b22}.

The hypothesis of the ML existence or Generalized Uncertainty Principle (GUP) has been done from a long time for conceptual as well as technical reasons. Such an important interest was motivated in the quantum gravity\cite{b79,b80,b81} and the string theory \cite{b82,b83,b84} which propose small corrections at Heisenberg Uncertainty Principle (HUP) implies non zero minimal uncertainty $  (\Delta x)_{min}= \hbar \sqrt{\alpha} $. However the concept of ML can be integrated on the study of the physical systems by considering the deformed canonical commutation relation
\begin{equation}
[\hat{X}, \hat{P}]=i h\left(1+\alpha^{2} \hat{P}^{2}\right),
\label{dccr}
\end{equation}

Where $ \alpha $ represents the ML parameter ( is very small positive parameter), this commutation relation leads to the uncertainty relation

\begin{equation}
(\Delta X)(\Delta P) \geq \frac{\hbar}{2}\left(1+\alpha(\Delta P)^{2}+\tau\right), 
\tau =\alpha\langle\widehat{P}\rangle^{2}
\label{ur}
\end{equation}

 As a matter of fact, the ML concept has been used in several quantum physical theories which have been favorable for the existence of a ML with magnitudes ranging from the Planck constant ($10^{-35}m$)\cite{b73,b74} to the order  $10^{-15}m$ \cite{b75,b76,b77}. In this optic, the ML concept has been introduced, for the first time, in nuclear structure through BMM in the pioneering work \cite{b20}.
\par In a  related context, the quantum Harmonic model potential with the contribution of sextic anharmonic terms has been used in several works\cite{b23,b24,b25,b26,b27,b28,b29,b30,b31}. Indeed, this class of potential takes an oscillator model role in some  quantum observables like spectra of hydrogen and bounded-solids\cite{b24}. Such a potential might be considered as a model potential for quark confinement in Quantum Chromodynamics\cite{b25} and its study continues to be a topic of current research interest\cite{b26,b27,b28,b29,b30,b31}, mainly in the field of nuclear structure due to its important role in shape coexistence and mixing phenomena from a geometrical perspective\cite{b31}.
It is well known that the sextic potential has the following form,
\begin{equation}
V(\xi_1,\xi_2,\xi_3;\beta)=\frac{\hbar^{2}}{2B}\left( \xi_1\beta^{2}+\xi_2\beta^{4}+\xi_3\beta^{6}\right) , \ \xi_3\neq 0,
\label{potg}
\end{equation}
which is considered as a quasi-exactly solvable phenomenological potential, i.e. a part of its energy spectrum can be exactly established if the parameters of the potential fill out certain restrictions. In order to adapt the same form of potential which was used previously in Refs\cite{b29,b30}, we choose the parameters $\xi_i$($i=1,2,3$) in Eq. \eref{potg} as follows :
\begin{eqnarray}
\xi_1=\left(b^{2}-4ac\right), \xi_2=2ab, \xi_3=a^{2} ,
\end{eqnarray}
where $a$, $b$ and $c$ are three parameters.
Also, it appears that for the particular value $a=0$ the sextic potential becomes the harmonic oscillator potential.
\par The present model is conventionally called X(3)-SML, and the plan of this work is as follows: Section II is devoted to the presentation of the model  which is divided in two subsections. The first one covers the theoretical results such as energy spectra and the corresponding wave functions  outside the ML formalism using quasi-exactly solvable method (QES) as in Refs\cite{b29,b30}, whereas the second deals with the presentation of  X(3)-SML model with the quantum perturbation method which is required by the study of the model in the  presence of the ML. In Section III the numerical applications are presented and the results are discussed from both theoretical and experimental point of view. Finally, Section IV contains our conclusions.

\section{\label{sec:level1}Theory of the model}

Theoretical background of ML formalism motivated by a Heisenberg algebra and implies a generalized uncertainty principle (GUP) has been considered recently in \cite{b86,b87}.
In the context of this formalism, the deformed canonical commutation relation \eref{dccr} has been generalized in order to have not only non-zero minimal uncertainty at the position, but also to insure that the position $ \hat{X}_{i} $ and momentum $ \hat{P}_{i} $ operators will be Self-adjoint operators \cite{b85}
this relation given by \cite{b86,b87}

\begin{equation}
\left[\hat{X}_{i}, \hat{P}_{i}\right]=i \hbar\left(\delta_{i j}+\alpha \hat{P}^{2} \delta_{i j}+\alpha^{\prime} \hat{P}_{i} \hat{P}_{j}\right), \left( \alpha,\alpha^{\prime}\right) > 0,
\end{equation}

where $\alpha^{\prime}$ is an additional parameter which is of order of $\alpha .$ In this case, the components of the momentum operator commute to one another
\begin{equation}
\left[\hat{P}_{i}, \hat{P}_{j}\right]=0
\end{equation}
So, Jacobi identity implies non-commutative algebra
\begin{equation}
\left[\hat{X}_{i}, \hat{X}_{j}\right]=i \hbar \frac{\left(2 \alpha-\alpha^{\prime}\right)+\left(2 \alpha+\alpha^{\prime}\right) \alpha \hat{P}^{2}}{1+\alpha \hat{P}^{2}}\left(\hat{P}_{i} \hat{X}_{j}-\hat{P}_{j} \hat{X}_{i}\right),
\end{equation} 
Supposing that $ \Delta P_{j} $ does not depend on j, the corresponding uncertainty relation is given by

\begin{equation}
\left(\Delta X_{i}\right)\left(\Delta P_{j}\right)\geq \frac{\hbar}{2} \delta_{i j}\left[1+\left(3 \alpha+\alpha^{\prime}\right)\left(\Delta P_{j}\right)^{2}+\lambda\right]
\label{ur}
\end{equation}
with
\begin{equation}
\lambda =\alpha \sum_{k=1}^{3}\left\langle P_{k}\right\rangle^{2}+\alpha^{\prime}\left\langle P_{i}\right\rangle^{2}
\end{equation}
The minimization of \eref{ur} in $ \Delta P_{j} $ leads to 
$\left(\Delta X_{i}\right)_{\min }=$ $\hbar \sqrt{3 \alpha+\alpha^{\prime}} .$
In the literature, several representations of the operators $  \hat{X}_{i} $ and $  \hat{P}_{i} $ have been used. Among these representations one can cite the momentum space representation \cite{b86}
\begin{equation}
\hat{X}_{i}=i \hbar\left[\left(1+\alpha p^{2}\right) \frac{\partial}{\partial p_{i}}+\alpha^{\prime} p_{i} p_{j} \frac{\partial}{\partial p_{j}}\right]+\eta p_{i}, \hat{P}_{i}=p_{i}
\end{equation}
and the position representation given by \cite{b88,b89}
\begin{equation}
\hat{X}_{i}=\hat{x}_{i}+\frac{\left(2 \alpha-\alpha^{\prime}\right)\left(\hat{p}^{2} \hat{x}_{i}+x_{i} \hat{p}^{2}\right)}{4}, \hat{P}_{i}=\hat{p}_{i}\left(1+\frac{\alpha^{\prime}}{2} \hat{p}^{2}\right)
\label{position}
\end{equation}
where $\hat{x}_{i}$ and $\hat{p}_{i}$ are the usual position and momentum operators respectively. Which obey the following relations $\left[ \hat{x}_{i}, \hat{p}_{j}\right]=i \hbar \delta_{i j}$ and $\hat{p}^{2}=\sum_{i=1}^{3}\hat{p}_{i}$.  in the case of $\alpha^{\prime}=2 \alpha$, the following canonical commutation $\left[\hat{X}_{i}, \hat{X}_{j}\right] .$ for the first order on $\alpha,$ vanishes. As a consequence Eq. \eref{position} reduces to.
\begin{equation}
\hat{X}_{i}=\hat{x}_{i}, \hat{P}_{i}=\left(1+\alpha \hat{p}^{2}\right) \hat{p}_{i}
\label{posrep}
\end{equation}
In addition, we can interpret $\hat{p}_{i}$ and $\hat{P}_{i}$ shown in Eq .\eref{posrep} as follows: $\hat{p}_{i}$ is the momentum operator at low energics and $\hat{P}_{i}$, is the momentum operator at high energies. Moreover, $\hat{p}$ is the magnitude of the $\hat{p}_{i}$ vector.

Here, we recall that, within the ML formalism, the collective quadrupole Hamiltonian of Bohr-Mottelson has the following form\cite{b20,b32,b33} :
\begin{equation}
H=-\frac{\hbar^2}{2B}\Delta+\frac{\alpha \hbar^4}{B}\Delta^2+V(\beta),
\label{Ham}
\end{equation}
with
\begin{equation}
\Delta=\frac{1}{\sqrt{g}} \partial_{i} \sqrt{g} g^{i j} \partial_{j}=\frac{1}{\beta^{2}} \frac{\partial}{\partial \beta} \beta^{2} \frac{\partial}{\partial \beta}-\frac{\Delta_{\Omega}}{\beta^{2}},
\end{equation}
 where $g$ is the determinant of the matrix $g_{i j}$ given by
\begin{equation}
g_{i j}=\left(\begin{array}{ccc}
3 \beta^{2} \sin ^{2} \theta & 0 & 0 \\
0 & 3 \beta^{2} & 0 \\
0 & 0 & 1
\end{array}\right),
\end{equation}
 and $\Delta_{\Omega}$ is the angular part of the Laplace operator
\begin{equation}
\Delta_{\Omega}= \frac{{\hat{I}}}{3}=\frac{1}{3}\left[ \frac{1}{\sin \theta} \frac{\partial}{\partial \theta} \sin \theta \frac{\partial}{\partial \theta}+\frac{1}{\sin ^{2} \theta} \frac{\partial^{2}}{\partial \phi^{2}}\right],
\end{equation}
checking
\begin{equation}
\Delta_{\Omega} Y_{L M}(\theta, \phi)=-\frac{L(L+1)}{3} Y_{L M}(\theta, \phi),
\end{equation}
where $L$ is the angular momentum quantum number, M is its projection and  $ Y_{LM}(\theta,\phi) $  are the well-known spherical harmonics. The corresponding deformed Schr$\ddot o$dinger  equation, to the first order in  $\alpha $, reads as 
\begin{equation}
\left[-\frac{\hbar^2}{2B}\Delta+\frac{\alpha \hbar^4}{2B}\Delta^2+V(\beta)-E \right ] \psi(\beta,\theta,\phi) =0.
\label{shr}
\end{equation}
The separation of variables is achieved by considering the auxiliary wave function as in \cite{b90} 
 \begin{equation}
 \psi(\beta,\theta,\phi )=\left[1+2\alpha\hbar^2 \Delta\right ]F(\beta)Y_{LM}(\theta,\phi) ,
 \end{equation}
which leads to the following
\begin{equation}
\fl \left[\frac{1}{\beta^2}\frac{d}{d\beta} \beta^2\frac{d}{d\beta} -{\frac {	 L   \left( L+1 \right) }{{3\beta}^{2}}}+\frac{2B}{\hbar^{2}}\left( \frac{E-V(\beta)}{1+4B\alpha(E-V(\beta))}\right)
\right]F( \beta ) =0.
\label{e7}
\end{equation}
By virtue of the complexity of this differential equation in its form and in order to find an approximate solution for it, we adapt an approximation of the term proportional to $(E-V(\beta))$ by expanding it in power series  of $\alpha $ as follows\cite{b33}:
\begin{eqnarray}
\left(1+4B\alpha(E-V(\beta))\right)^{-1}\simeq 1-4B\alpha(E-V(\beta)),
\end{eqnarray}
which is a valid approximation owing to the smallness of the parameter $\alpha$.
However the above differential equation was solved exactly with an infinite square well like potential within the standard method\cite{b20}, and approximately with a scaling Davidson potential\cite{b32} by means conjointly of Asymptotic Iteration Method (AIM)\cite{b35} and a quantum perturbation method. Also the Coulomb and Hulthen potentials were also studied\cite{b33} in this context.

\subsection{Solutions of $\beta$-equation for $\alpha=0$ within QES}
Our task now is to find eigenvalues and eigenfunctions of the Hamiltonian \eref{Ham} with the sextic potential in the case of $\alpha=0$. For this purpose, it is preferable to write equation \eref{e7} in a Schr$\ddot o$dinger picture. This is realized easily by changing the unperturbed wave function  $F^{(0)}( \beta )$ by $\frac{\xi( \beta )}{ \beta}$ :
\begin{equation}
\left [-{\frac {d^{2}}{d{\beta}^{2}}}  +{\frac {L(L+1)}{{3\beta}^{2}}}+ v \left( \beta \right)\right ]\xi \left( 	\beta \right) =\varepsilon \xi\left( 
\beta \right),
\label{e10}
\end{equation}
with,
\begin{equation}
\varepsilon=\frac{2B}{\hbar^{2}}E\\    
,	 v \left( \beta \right)=\frac{2B}{\hbar^{2}}  V \left( \beta \right).\\
\end{equation}
The Hamiltonian of the sextic oscillator with a centrifugal barrier potential has the following expression\cite{b26}:
\begin{equation}
\fl H_{\beta}=-{\frac {d^{2}}{d{\beta}^{2}}} +{\frac {\left( 2\,s-\frac{1}{2} \right)  \left( 2\,s-\frac{3}{2} \right) }{{\beta}^{2}}}+ \left[b^{2}-4a(s+\frac{1}{2}+M^{\prime})\right] \beta^{2} + 2ab\beta^{4}+a^{2}\beta^{6},
\label{e11}
\end{equation}
where $s$ defines the centrifugal term of the effective potential and usually depends on some quantum numbers such as total angular momentum or its projections. $M^{\prime}$ is a natural number which fixes the size of the finite matrix for which exact solutions can be found. Thus, for given $s$ and $M^{\prime}$, one will have $M^{\prime}+$1 exact solutions, and therefore the exact solvability order of the corresponding potential is $M^{\prime}+1$. This can be  readily checked if we take as an appropriate ansatz the function $\xi(\beta)=P^{(M^{\prime})}(\beta^2)\beta^{2s-\frac{1}{2}}e^{-\frac{a\beta^4}{4}-\frac{b\beta^2}{2}}$ in  the  collective eigenvalues equation $H_{\beta}\xi(\beta)=\varepsilon\xi(\beta)$, where $P^{(M^{\prime})}(\beta^2)$ is a polynomial in $\beta$ of degree $M^{\prime}$. So, we obtain\cite{b21,b29} :
\begin{equation}
Q_{\beta}P^{(M^{\prime})}(\beta^2)=\lambda P^{(M^{\prime})}(\beta^2),
\label{Matrixeq}
\end{equation}
with 
\begin{equation}
\fl Q_{\beta}=-\left(\frac{\partial^2}{\partial\beta^2}+\frac{4s-1}{\beta}\frac{\partial}{\partial\beta}\right)+2b\beta\frac{\partial}{\partial\beta}+2a\beta^2\left(\beta\frac{\partial}{\partial\beta}-2M^{\prime}\right),
\end{equation}
 $ \lambda = \varepsilon-4bs $ and $ \ \ M^{\prime}=0,1,2,3,\cdots $

The connection of the differential equation in the presence of sextic potential with that corresponding to the Hamiltonian \eref{e11} is made at the level of the centrifugal barrier by considering the relations\cite{b29}:
\begin{equation}
\left(2s-\frac{1}{2}\right)(2s-\frac{3}{2})=\frac{L(L+1)}{3}.
\label{e12}
\end{equation}
On the other hand, we can see that, the sextic potential depends on two parameters, $a$ and $b$, and on an integer $M^{\prime}$, but also on the quantum
number $L$ through $s$. Seemingly, the potential is state dependent because $s$ depends on $L$, but, as it will be shown, $M^{\prime}$ can be used instead to keep the potential independent on $L$ by imposing that:
\begin{equation}
s+\frac{1}{2}+M^{\prime}= const =c. 
\label{e13}
\end{equation}
By using Eq. \eref{e12} the expression of s can be obtained as a function of L, 
\begin{equation}
s(L)=\frac{1}{2}\left [1+\sqrt{\frac{L(L+1)}{3}}+\frac{1}{4}\right].
\end{equation}
For $L=0$ and 2 the expression
of s  reduces to a simpler form:
\begin{equation}
s^{'}\left(L\right )=\frac{L+3}{4}
\label{sLp}
\end{equation}
The condition \eref{e13} becomes
\begin{equation}
c=M^{\prime}+\frac{L}{4}+\frac{5}{4},
\end{equation}
which is constant if $ M^{\prime} $ decreases by one unit while L increases by four units\cite{b29} :
\begin{eqnarray}
&(M^{\prime},L):(k,0),(k-1,4),(k-2,8)....\Rightarrow k+\frac{5}{4}=c^{(k)}_{0},\\
\label{ee19}
&(M^{\prime},L):(k,2),(k-1,6),(k-2,10)....\Rightarrow k+\frac{7}{4}=c^{(k)}_{2}.
\label{ee20}
\end{eqnarray}
Here by k is denoted the maximum value of $ M^{\prime} $. It is also called the quasi-exact solvability order giving the number of solutions which are exactly determined.

The sextic potential equation \eref{e10} can be more simplified by reducing the number of parameters through the change of variable $\beta=y a^{-1 / 4}$ and adopting the notations $\varrho=b / \sqrt{a}$ and $\varepsilon_{y}=\varepsilon / \sqrt{a}$ :
\begin{equation}
\left[-\frac{d^{2}}{d y^{2}}+\frac{L(L+1)}{3 y^{2}}+v_{m}^{(k)}(y)\right] \eta^{(0)}(y)=\varepsilon_{y} \eta^{(0)}(y)
\label{eq35}
\end{equation}
where,
\begin{equation}
v_{m}^{(k)}(y)=(\varrho^{2}-4c_{m}^{(k)})y^{2}+2\varrho y^{4}+y^{6}+u_{m}^{(k)}(\varrho ) \ \,  m=0,2.
\label{eq36}
\end{equation}

is the improved potential energy checking that the constraint of the constant potential is exactly satisfied.

Here, we present explicitly the different constant $ u_{m}^{(k)}(\varrho) $ as  in Ref.\cite{b26}. The slight ambiguity of the potential expression was handled by selecting the constant terms $u_{0}^{(k)}(\varrho)$ and $u_{2}^{(k)}(\varrho)$ such that the minima of the two potentials are set at the same energy. This requirement can be fulfilled by setting $u_{0}^{(k)}(\varrho)=0$ and

\begin{equation}
\fl u_{2}^{(k)}(\varrho)=\left\{\begin{array}{ll}\left(\varrho^{2}-4 c_{0}^{(k)} \right)\left(y_{0,0}^{(k)}\right)^{2}-\left(\varrho^{2}-4 c_{2}^{(k)}\right)\left(y_{0,2}^{(k)}\right)^{2}\\+2\varrho\left[\left(y_{0,0}^{(k)}\right)^{4}-\left(y_{0,2}^{(k)}\right)^{4}\right]  +\left[\left(y_{0,0}^{(k)}\right)^{6}-\left(y_{0,2}^{(k)}\right)^{6}\right] &{ if } \quad \varrho^{2}<4 c_{0}^{(k)} \\ -\left(\varrho^{2}-4 c_{2}^{(k)}\right)\left(y_{0,2}^{(k)}\right)^{2}-2 b\left(y_{0,2}^{(k)}\right)^{4}-\left(y_{0,2}^{(k)}\right)^{6}&{ if } \quad 4 c_{0}^{(k)}<\varrho^{2}<4 c_{2}^{(k)} \\ 0 &  { if } \quad \varrho^{2}>4 c_{2}^{(k)}\end{array}\right.
\end{equation}

where 
\begin{equation}
(y_{0,m}^{(k)})^{2}=\frac{1}{3}\left(-2\varrho+\sqrt{\varrho^{2}+12c_{m}^{(k)}} \right ), m=0,2.
\end{equation}
are the absolute potential minima of the even $ \frac{L}{2} $ and odd  $ \frac{L}{2} $ potentials. For $ \varrho^{2}<4 c_{0}^{(k)} $, one has $ y_{0,m}^{(k)}> 0 $. Instead for \quad $ 4 c_{0}^{(k)}<\varrho^{2}<4 c_{2}^{(k)} $,  $ y_{0,0}^{(k)}= 0 $ and  $ y_{0,2}^{(k)}> 0 $. Finally, for $ \varrho^{2}>4 a c_{2}^{(k)} $, both absolute minima become spherical. All these constants are taken into consideration in the fit of experimental data .
Therefore, according to \cite{b29}, by considering the ansatz function:
\begin{equation}
\eta^{(0)}(y) = \eta_{n, L}^{(0)}(y)=N_{n, L} P_{n, L}^{(M^{\prime})}\left(y^{2}\right) y^{\frac{L}{2}+1} e^{-\frac{y^{4}}{4} -\frac{\varrho y^{2}}{2} }, n=0,1,2, 
\end{equation}

where $ N_{n, L} $ are the normalisation factor,
while $ P_{n, L}^{(M^{\prime})}\left(y^{2}\right) $ are polynomials in $ y^{2} $ of $ M^{\prime} $ order. Eq. \eref{eq35} with potential \eref{eq36} is reduced to the equation
\begin{equation}
\fl{\left[-\left(\frac{d^{2}}{d y^{2}}+\frac{4 s{'}-1}{y} \frac{d}{d y}\right)+2 \varrho \frac{d}{d y}\right] P_{n, L}^{(M^{\prime})}\left(y^{2}\right)}+2 y^{2}\left(y \frac{d}{d y}-2 M^{\prime}\right) P_{n, L}^{(M^{\prime})}\left(y^{2}\right) =\lambda P_{n, L}^{(M^{\prime})}\left(y^{2}\right)
\end{equation}

Taking into account the above considerations and matrix form of \eref{Matrixeq}, the eigenvalues are obtained as
\begin{equation}
\lambda =\lambda^{(k)}_{n,L}(\varrho)=\varepsilon_{y}-4bs{'}-u^{(k)}_{m}(\varrho)-\frac{1}{\left<y^{2}\right>_{n,L}}\frac{L}{6}\left( \frac{L}{2}-1\right),
\label{eq27}
\end{equation}
where the last term, containing the mean value of $ y^{2} $ given by
\begin{equation}
\left<y^{2}\right>_{n,L}=\left<\eta^{(0)}_{n,L}( y )\left|y^{2}\right|\eta^{(0)}_{n,L}( y )\right> =\int_{0}^{\infty} \eta_{n, L}^{(0)}(y) y^{2} \eta_{n, L}^{(0)}(y) d y,	
\end{equation}
 was extracted from the centrifugal
term of \eref{e10} such that $s(L)$ could be approximated by $ s{'}(L) $ \eref{sLp} for $ L\geq 4 $.
Finally, the total energy of the system in the case of  $\alpha=0$ is easily deduced from  \eref{eq27}:
\begin{equation}
\fl E^{(0)}_{n,L}=E =\frac{\hbar^{2}\sqrt{a}}{2B}\left [\lambda^{(k)}_{n,L}\left (\varrho\right )+\varrho(L+3)+u^{(k)}_{m}(\varrho)+\frac{1}{\left<y^{2}\right>_{n,L}}\frac{L}{6}\left( \frac{L}{2}-1\right)\right ].
\label{e26}
\end{equation}

The total energy \eref{e26} depends on two quantum numbers, the order of the wave function's zero $n$ and the total angular momentum $L,$ on a fixed integer $k,$ on a scale factor $ \frac{\hbar^{2}\sqrt{a}}{2 B} $  and on a free parameters $\varrho$.
 In the present paper, the  method is
adopted with n = 0, n = 1 and n = 2 corresponding to the ground band, the first and
second $ \beta $ bands, respectively.
 
It should be noted here that the model presented in this subsection is called X(3)-Sextic and its solutions were carried out in \cite{b29}. We can therefore say that the results obtained here are only a summary of the work which  will obviously help us in our current work. So, for more details, we refer the reader to \cite{b26,b27,b28,b29,b30}.

\subsection{Correction to the energy spectrum of system by QPM}

Before the treatment of perturbation term, it is important to mention that Eq. \eref{shr} was previously solved for an ISWP \cite{b20}, Davidson potential (D) \cite{b32}, Hulthèn and  Coulomb potentials \cite{b33}. All these potentials are presented in \Tref{tabmodels} in order to be compared with our model.

\begin{table}[ht!]
	\caption{The potentials in the $\beta$ variable and the $ \gamma $
		rigidity values for the relevant $  \gamma$-rigid  solutions in the presence of the ML.}
	\vspace{0.3cm}
\begin{center}
		\footnotesize
		\begin{tabular}{p{4.5cm}*{3}{@{\hskip.8mm}c@{\hskip.8mm}}}
			\br 
			 Solution  & $ \beta $  potential  & $ \gamma $& Ref \\
			\mr
			 X(3)-MLF  & {0,  if $  \beta \leq \beta_{\omega} $,} & {$ 0^{\circ} $}& \cite{b20} \\
			{} & {$ \infty$,  if  $\beta>\beta_{\omega} $} & {} & \\
			 X(3)-D-ML  & {$ a \beta^{2}+\frac{b}{\beta^{2}} $} & {$ 0^{\circ} $} &\cite{b32}\\
			X(3)-H-ML  & {$ \frac{e^{-\delta \beta}}{e^{-\delta \beta}-1} $} & {$ 0^{\circ} $} &\cite{b33}\\
			 X(3)-C-ML  &   {$ \frac{c}{ \beta } $}  &  {$ 0^{\circ} $} & \cite{b33}\\
			\br	
		\end{tabular}
 \end{center}
	\label{tabmodels}
\end{table}
\normalsize
In the perturbation theory, the corrected energy spectrum can be written as\cite{b32}:
\begin{equation}
E^{C o r r}_{n,L}=E^{(0)}_{n,L}+\Delta E_{n,L},	
\label{e27}	
\end{equation}
where $E^{(0)}_{n,L}$ are the unperturbed levels corresponding to the eigenfunctions solutions of Schr$\ddot o$dinger equation for $\alpha=0$ and  $\Delta E_{n,L}$ is the correction induced by the ML, given by:
\begin{equation}
\Delta E_{n,L}=\alpha\frac{\hbar^{4}}{B}\left<F^{(0)}_{n,L}( \beta )\left|\Delta^{2}\right|F^{(0)}_{n,L}( \beta )\right>,	
\end{equation}
 which is the mean value of the perturbation taken with respect to the unperturbed eigenstate $ F^{(0)}_{n,L}( \beta ) $.
It can also be expressed as,
\begin{equation}
\fl \Delta E_{n, L}=4 B \alpha\left[\left(E_{n, L}^{(0)}\right)^{2}-2 E_{n, L}^{(0)}\left\langle F_{n, L}^{(0)}|V(\beta)| F_{n, L}^{(0)}\right\rangle+\left\langle F_{n, L}^{(0)}\left|(V(\beta))^{2}\right| F_{n, L}^{(0)}\right\rangle\right].
\label{e29}
\end{equation}
After substituting the sextic potential\eref{eq36} into \eref{e29} and using the change of variable $\beta=y a^{-1 / 4}$, one obtains	
\begin{eqnarray}
\fl \Delta E_{n,L}=&\kappa \sqrt{a}\frac{\hbar^{4}}{B}\bigg[ \overline{{y}^{12}}+4 \varrho\overline{{y}^{10}}+ 2 \left(2 \varrho^{2}+A_{m}^{(k)} \right) \overline{{y}^{8}}+ 2 \left(2 \varrho^{2}A_{m}^{(k)}-\varsigma^{(0)}_{n,L} \right)  \overline{{y}^{6}}\nonumber\\&+ \left((A_{m}^{(k)})^{2}-4 \varrho\varsigma^{(0)}_{n,L} \right)  \overline{{y}^{4}}- 2 A_{m}^{(k)}\varsigma^{(0)}_{n,L}   \overline{{y}^{2}}+ \left({u^{(k)}_{m}(\varrho)}-\varsigma^{(0)}_{n,L} \right)\bigg],
\end{eqnarray}
	
where
$A_{m}^{(k)} =\left(\varrho^{2}-4\,c_{m}^{(k)}\right )$, $ \varsigma^{(0)}_{n,L}=\lambda^{(k)}_{n,L}\left (\varrho\right )+\varrho(L+3)+\frac{1}{\left<y^{2}\right>_{n,L}}\frac{L}{6}\left( \frac{L}{2}-1\right) $, $ \kappa =\sqrt{a}\alpha $ is the new ML parameter  and $\overline{y^{t}} (t=2, 4,6,8,12)$ are the mean value of  $y^{t}$ given by:

\begin{equation}
\overline{y^{t}}=\int_{0}^{\infty} \eta_{n, L}^{(0)}(y) y^{t} \eta_{n, L}^{(0)}(y) d y
\end{equation}

 calculated as in Refs \cite{b71} \cite{b72}.

As the energy levels in quantum mechanical problems, the wave function is an important mathematical object. So, by using the same method, the corrected wave function in first-order approximation can be calculated by \cite{b32}: 
\begin{equation}
\fl \eta_{n, L}^{C o r r}(y)=\eta^{(0)}_{n,L}(y)+\sum_{k \neq n}\left[\frac{\int_{0}^{\infty} \eta^{(0)}_{k,L}(y)  \vartheta\left(n,\varrho,\kappa, \varsigma_{n, L}^{(0)},A_{m}^{(k)}\right)  \eta^{(0)}_{n,L}(y) d y}{\varsigma_{n, L}^{(0)}-\varsigma_{k, L}^{(0)}}\right]  \eta^{(0)}_{k,L}(y)
,
\label{fctcor}
\end{equation}
where

\begin{eqnarray}
\fl \vartheta\left(n,\varrho,\kappa, \varsigma_{n, L}^{(0)},A_{m}^{(k)}\right)=&2\kappa \hbar^{2}\bigg[{y}^{12}+4 \varrho {y}^{10}+ 2 \left(2 \varrho^{2}+A_{m}^{(k)} \right) {y}^{8}+ 2 \left(2 \varrho^{2}A_{m}^{(k)}-\varsigma^{(0)}_{n,L} \right) {y}^{6}\nonumber\\&+ \left((A_{m}^{(k)})^{2}-4 \varrho\varsigma^{(0)}_{n,L} \right) {y}^{4}- 2 A_{m}^{(k)}\varsigma^{(0)}_{n,L}{y}^{2}+ \left({u^{(k)}_{m}(\varrho)}-\varsigma^{(0)}_{n,L} \right)\bigg].
\end{eqnarray}

In the same context, the probability density distribution can be also evaluated via the following formula\cite{b32} :
\begin{equation}
\rho^{Corr}_{n,L}( \beta )= \beta^{2}\left|F^{Corr}_{n,L}( \beta )\right|^{2}= a^{-1/2}y^{2}\left|\eta^{Corr}_{n,L}(y)\right|^{2}.
\end{equation}
\subsection{E2 transition probabilities}

Having the corrected wave functions  $ \eta_{n, L}^{C o r r}(y) $, we can also calculate the expression of the reduced E2 transition probabilities.

In the general case the quadrupole operator is
\begin{equation}
T_{\mu}^{(E 2)}=t \beta\left[D_{\mu, 0}^{2}(\Omega) \cos \gamma+\frac{1}{\sqrt{2}}\left[D_{\mu, 2}^{2 *}(\Omega)+D_{\mu,-2}^{2 \cdot}(\Omega)\right] \sin \gamma\right],
\end{equation}
where $\Omega$ denotes the Euler angles and $t$ is a scale factor. For $\gamma=0$ the quadrupole operator
becomes \cite{b5}:
\begin{equation}
T_{\mu}^{(E 2)}=t \beta \sqrt{\frac{4 \pi}{5}} Y_{2 \mu}(\theta, \phi).
\label{opper}
\end{equation}
$B(E 2)$ transition rates
\begin{equation}
B\left(E 2 ; n L \rightarrow n^{\prime} L^{\prime}\right)=\frac{1}{2 L+1}\left|\left\langle n^{\prime} L^{\prime}|| T^{(E 2)}|| n L\right\rangle\right|^{2}
\label{BE2}
\end{equation}
are calculated using the wave functions of \Eref{fctcor} and the volume element $d \tau=\beta^{2} \sin \theta d \beta d \theta d \phi = a^{-3/4} y^{2} \sin \theta d y d \theta d \phi$. 
Therefore, the final expression of the reduced E2 transition probabilities, normalized to the
transition from the first excited state to the ground state, is given by :\\
\begin{equation}
B(E_{2}) =	T_{n,L,n^{'},L^{'}}=\frac{B\left(E2; n; L\rightarrow n^{'},L^{'}\right) }{B\left(E2; 0; 2\rightarrow 0,0\right )}=\left( \frac{C^{L2L^{'}}_{0,0,0} I^{Corr}_{n,L,n^{'},L^{'}}}{C^{220}_{0,0,0} I^{Corr}_{0,2,0,0}}\right) ^2,
\label{transition rates} 
\label{eq39}
\end{equation}

where the radial matrix element $I^{Corr}_{n,L,n^{'},L^{'}} $ can be given either in $ \beta $ or $ y $ variable:

\begin{eqnarray}
I^{Corr}_{n,L,n^{'},L^{'}}=&\int_{0}^{+\infty}	F^{Corr}_{n,L}( \beta ) \beta F^{Corr}_{n^{'},L^{'}}( \beta ) \beta^{2} \mathrm{d}\beta\nonumber\\&= a^{-1/4} \int_{0}^{+\infty}	\eta^{Corr}_{n,L}( y ) y \eta^{Corr}_{n^{'},L^{'}}( \beta ) \mathrm{d}y.
\end{eqnarray}

 calculated as in Refs \cite{b71} \cite{b72}.

 \section{\label{sec:level15}Numerical results and discussion}

\subsection{\label{sec:level2}Theoretical aspects of the ML for X(3)-Sextic}
The model established in this work, namely: X(3)-SML, is adequate for description of $ \gamma$-rigid nuclei for which the parameter $  \gamma $ is fixed to $ \gamma=0 $.
Basically, the energy levels of the ground state (g.s) band as well as those of the two $ \beta $ bands are characterized by the principal quantum number $ n=0 $, $ n=1 $ and $ n=2 $, respectively.
In this work, the discussion will be restricted to the case k = 2 for which, as can be deduced from \eref{ee19}, the properties of the most representative low-lying states of the g.s band (up to $ 10^{+} $), the first $ \beta $ band (up to $ 6^{+} $) and the second  $ \beta $ band (up to $ 2^{+} $) are  reproduced. At this point, the index k will be dropped as it has a fixed value. If necessary, results for $  k > 2 $ can be considered without any problem.

\bgroup
\begin{figure}[!htbp]
	{\includegraphics[scale=0.62]{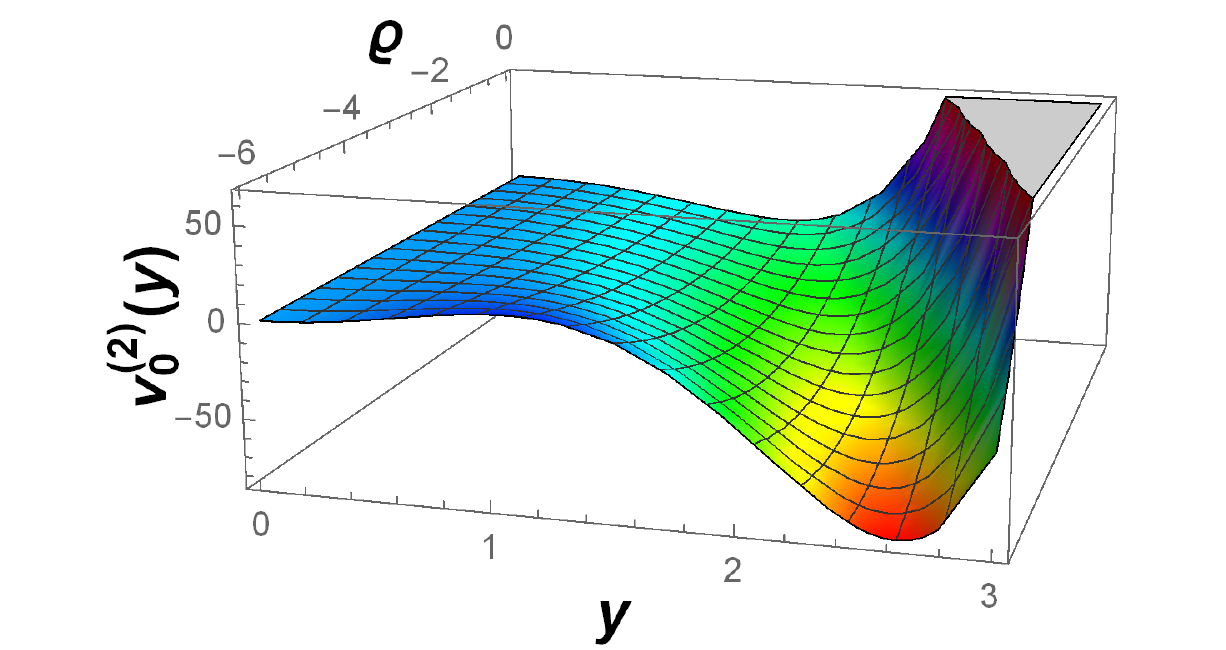}}{}{\includegraphics[scale=0.62]{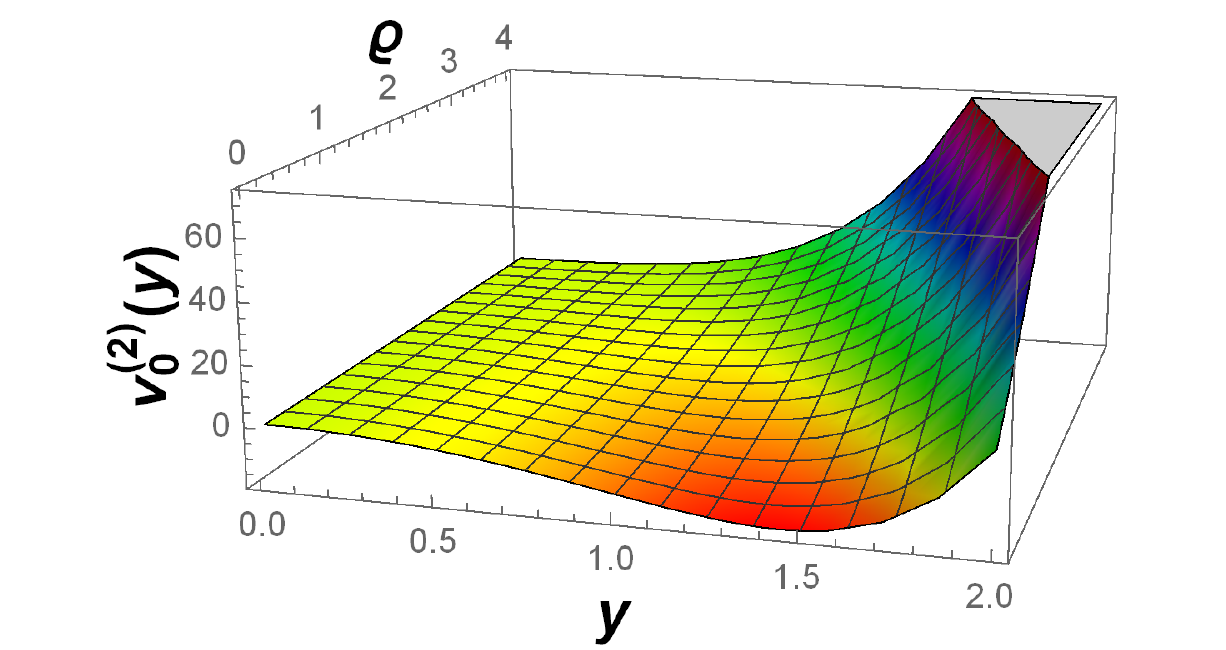}}{}
	\makeatother 
	\caption{{The shape evolution of the  sextic potential $ v^{(2)}_{0} (y) $, given by \eref{eq36}, as a function of the  parameter $\varrho$  for $ c^{(2)}_{0} = \frac{13}{4} $ and $ u^{(2)}_{0} = 0 $.}}
	\label{pot1}
\end{figure}
\egroup

Before discussing in more details the obtained numerical  results in our study, it is very worthwhile to recall here that the energy potential in \eref{eq36} is very appropriate to describe shape phase transitions and critical points due to the fact that, depending on its free parameter $\varrho$,  it can have a spherical minimum, a deformed one, a flat form or two simultaneous minima separated by a barrier ( maximum) as can be seen from \fref{pot1} which is an elegant representation in 3D.\\
The energy  \eref{e27}, up to scaling factor($ (\sqrt{a}\hbar^2)/2B $), depends only on two free parameters $ \varrho $ and $ \kappa $ and on two quantum numbers n and L. For numerical applications to experimental data, it is useful to normalize the energy \eref{e27} to the first excited energy
of the ground band:
\begin{equation}
R_{n, L}=\frac{E^{C o r r}_{n_, L}-E^{C o r r}_{0,0}}{E^{C o r r}_{0,2}-E^{C o r r}_{0,0}}.
\label{Rnl}
\end{equation}
By this normalization of the energies, the scaling factor ($ (\sqrt{a}\hbar^2)/2B $) is removed.
For the sake of revealing the ML effect on energy spectrum, in the context of $\gamma$-rigid nuclei, especially in the critical point, in this subsection we analyze the plot of energy ratios   R$_{nL}$   given in \Eref{Rnl} as a function of $\varrho$ for $ \kappa  =0$ , $ 0.001 $, $  0.01 $ and $  0.1 $ in \fref{energy rates}. From this latter, we can point out some useful remarks: in the absence of ML  ($\kappa=0$) , for higher values of the parameter $ \varrho $ ($\varrho > 6 $), $ X(3)-$SML  reproduces the results of the $ \gamma$-rigid prolate harmonic vibrator called $X(3)-\beta^{2} $ \cite{b36}, while for $\varrho$ close to zero a $  \gamma$-rigid prolate anharmonic vibrator shows up and for $ \varrho \in \left[ 2\sqrt{c^{(2)}_{0}}, 2\sqrt{c^{(2)}_{2}}\right] $ ( between the gray vertical lines ), a critical point of a first order shape phase transition between spherical and deformed shape occurs.
\bgroup
\begin{figure}[!htbp]
	{\includegraphics[scale=0.60]{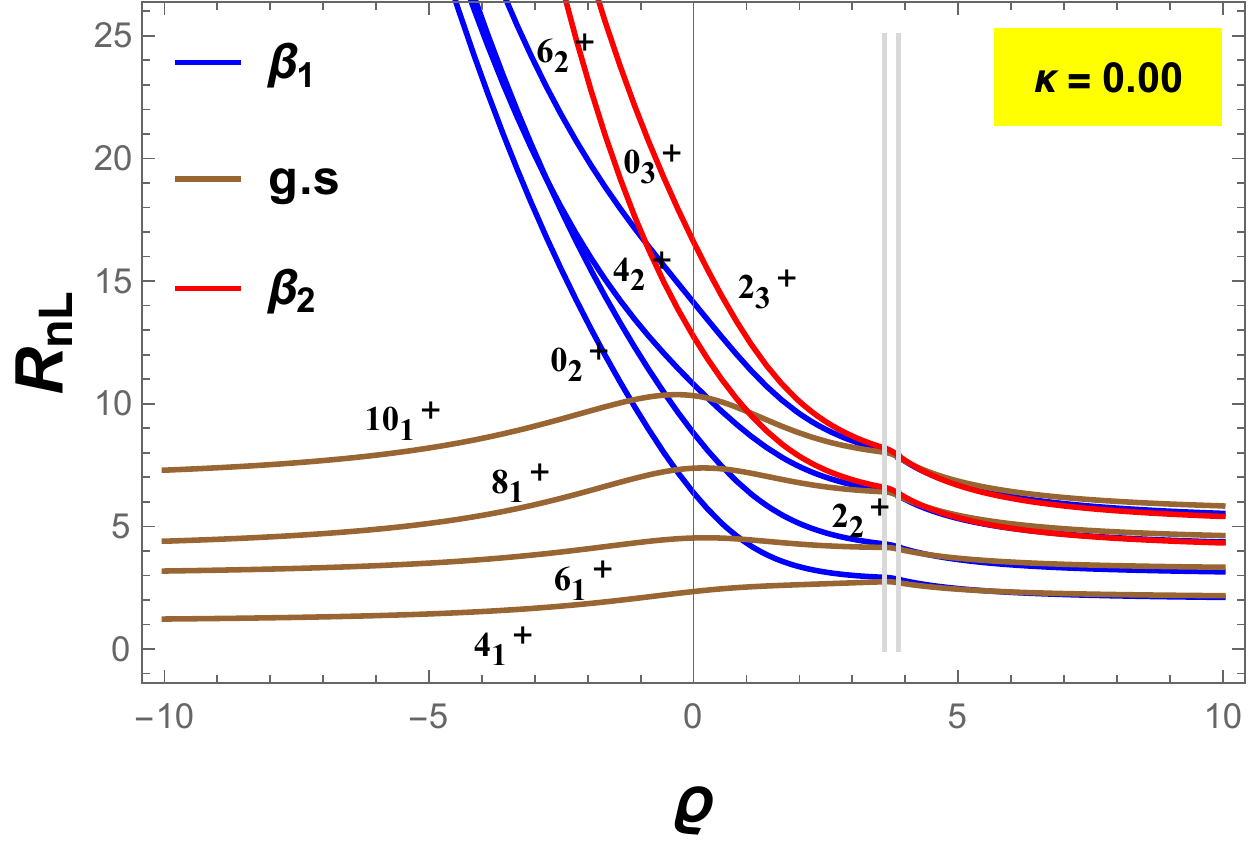}}{}	{\includegraphics[scale=0.60]{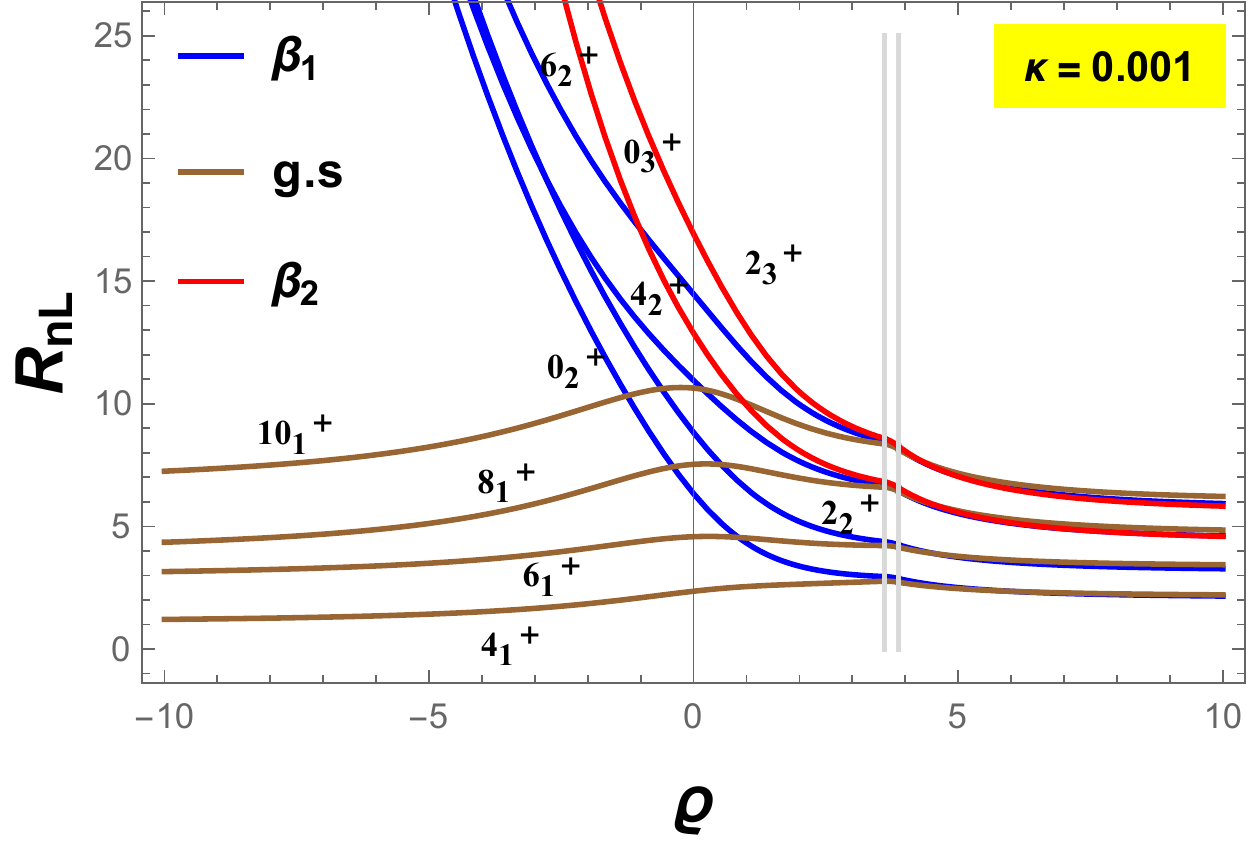}}{}\\
	{\includegraphics[scale=0.60]{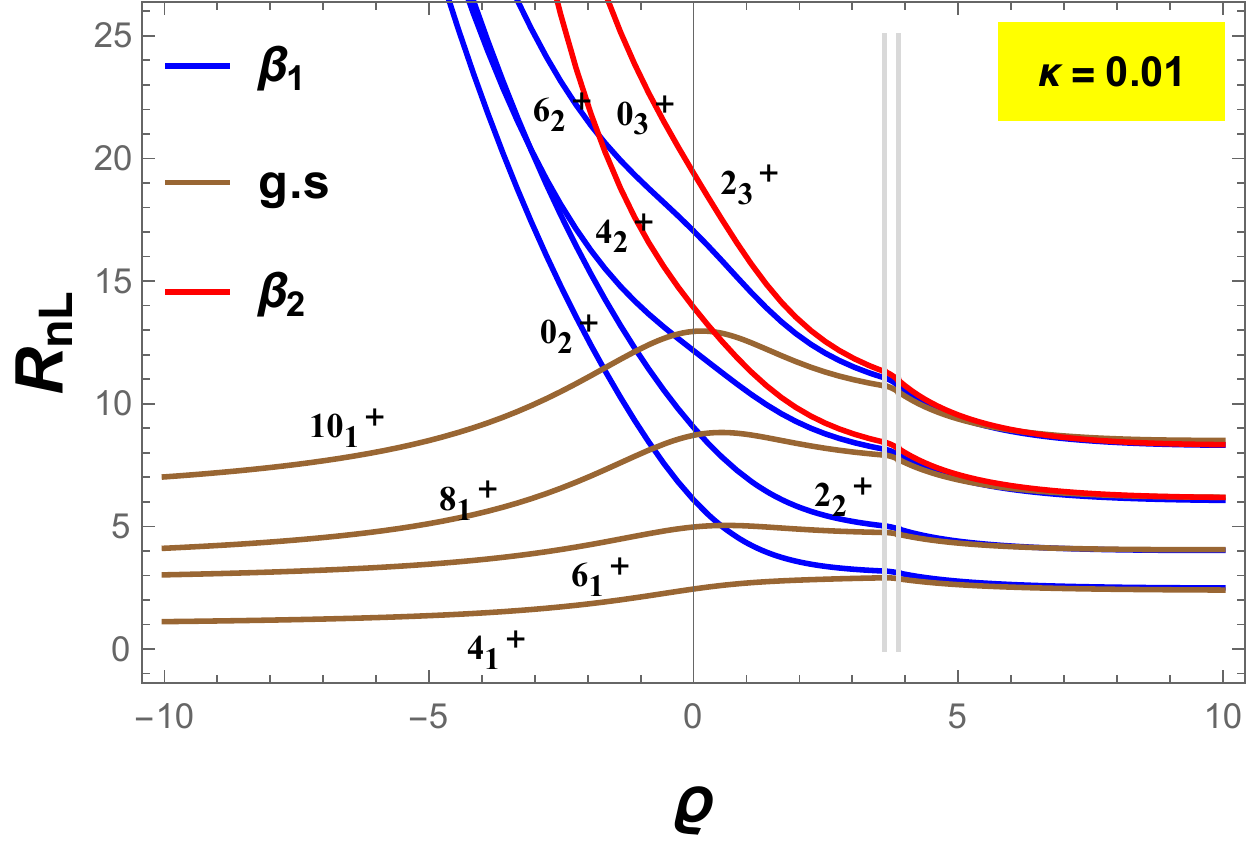}}{}	{\includegraphics[scale=0.60]{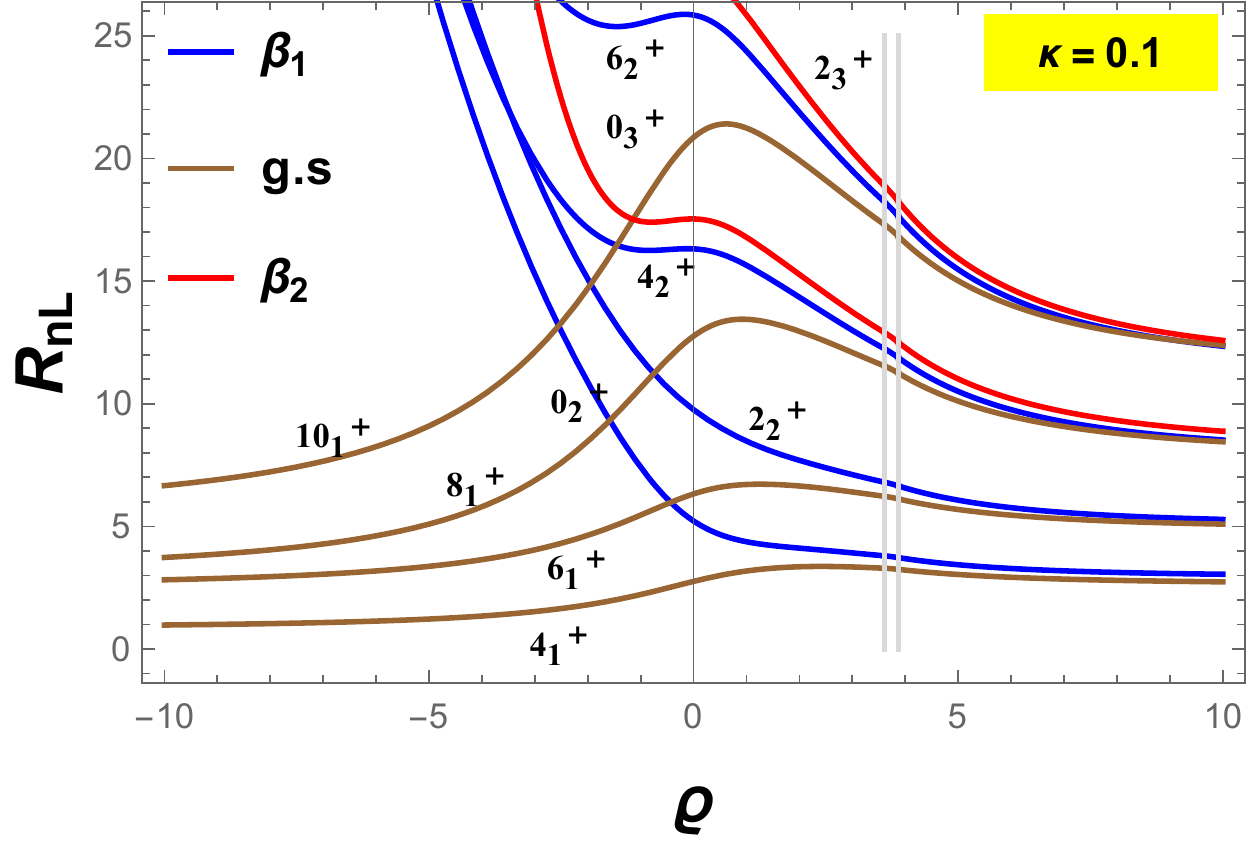}}{}
	\makeatother 
	\caption{{The energy spectra of the ground band and of the first two $  \beta $ bands,
			given by \eref{Rnl}, are plotted as a function of the  parameter $\varrho $ for  $\kappa =0 $, $ 0.001 $, $  0.01 $ , $  0.1 $ and  k = 2 . The energy lines are indexed by $ L^{+}_{n+1} $, where $ n = 0 $ for the
			ground band, $ n = 1 $ for the first $ \beta $ band and $ n = 2 $ for the second $ \beta $ band.
			.}}
	\label{energy rates}
\end{figure}
\egroup
 Besides, in the absence of the ML, the present model reproduces exactly the same behavior of the model X(3)-S \cite{b29}, namely: for values of $\varrho$ undergoing the constraint $\varrho \ll $ -6 , the $ \beta $ band energies go to infinity, while the g.s band energies keep finite values because the energy ratios do not depend on $\varrho$.
 While in the presence of the ML, for $ \varrho<-2\sqrt{c^{(2)}_{0}}$, the g.s band energies present local maximum which are slightly shifted forward on the right side in respect to the critical point region when $\kappa$ is raised, also the size of local maximum  increases with $\kappa$, concerning the two $ \beta $ band they go to infinity rapidly when $\kappa$ is increased, also, for  ($ \kappa  > 0.03 $),  these latters acquire additional nodes in the neighbourhood of  ($\varrho=0 $).  

Therefore, as we can observe from the \fref{energy rates}, the ML increases slightly both the energy ratios and the spacing between two successive levels.
In the region $\varrho \in \left[ \varrho_{min},- 2\sqrt{c^{(2)}_{0}}\right] $, where  $\varrho_{min} \ll$ -6, as can be seen from \fref{pot1}, spherical and deformed minima appear separated by a high barrier. On the other hand, in the interval $\varrho\in\left[ 2\sqrt{c^{(2)}_{0}}, 2\sqrt{c^{(2)}_{2}}\right] $, the shape of the energy potential in this case takes a form similar to that of an infinite square well, which plays an essential role in identifying critical points. This is also supported by the fact that the first derivative of the energy in
the critical point $\varrho_{c}\in \left[ 2\sqrt{c^{(2)}_{0}}, 2\sqrt{c^{(2)}_{2}}\right] $ has a discontinuity. Physically speaking, these famous results can be interpreted as follows: it is that a first order shape phase transition occurs between two $ \gamma $-rigid prolate vibrators: the first one is a harmonic type and the second one is an anharmonic kind.

\bgroup
\begin{figure}[!htbp]
	
	{\includegraphics[scale=0.52]{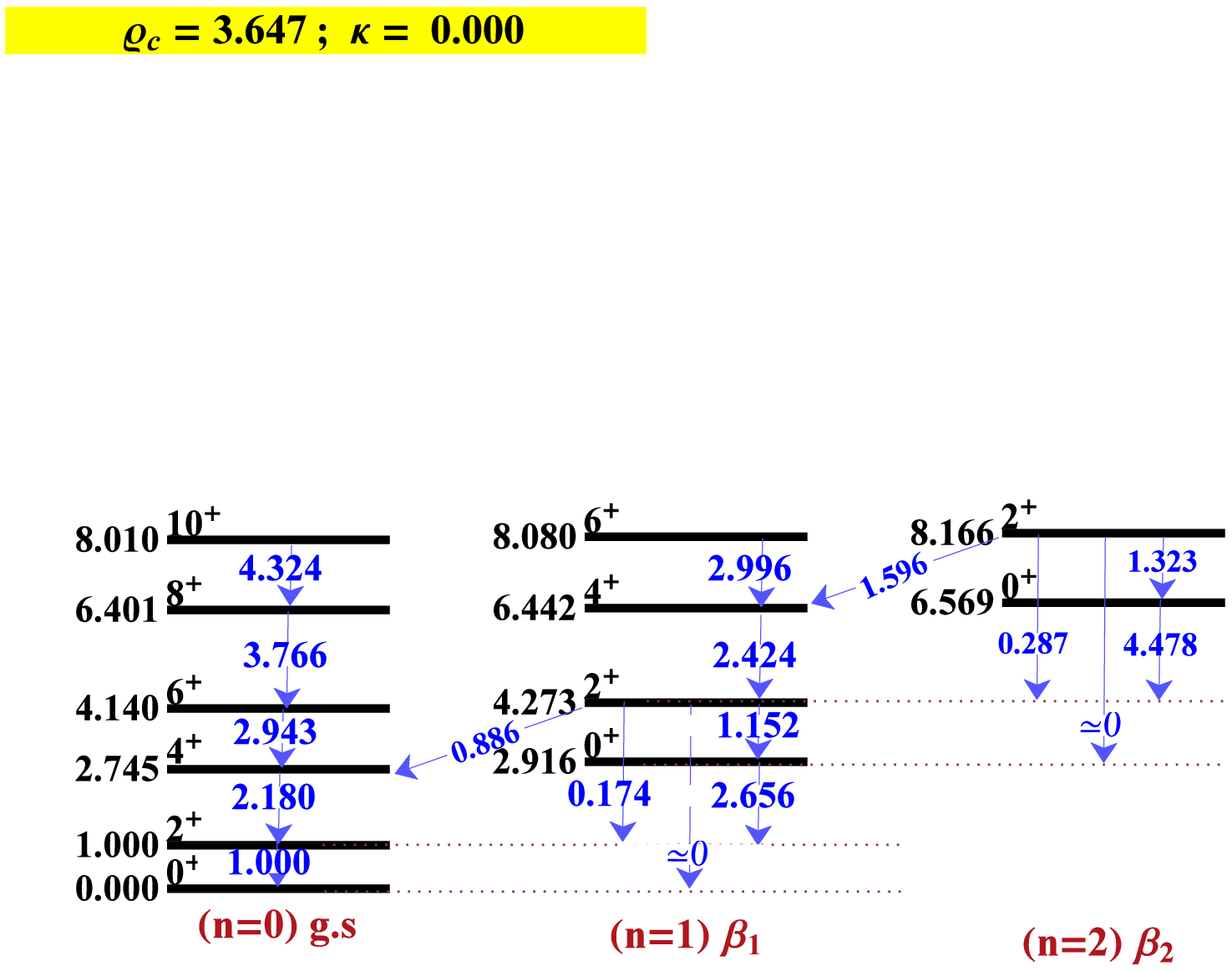}}{}{\includegraphics[scale=0.52]{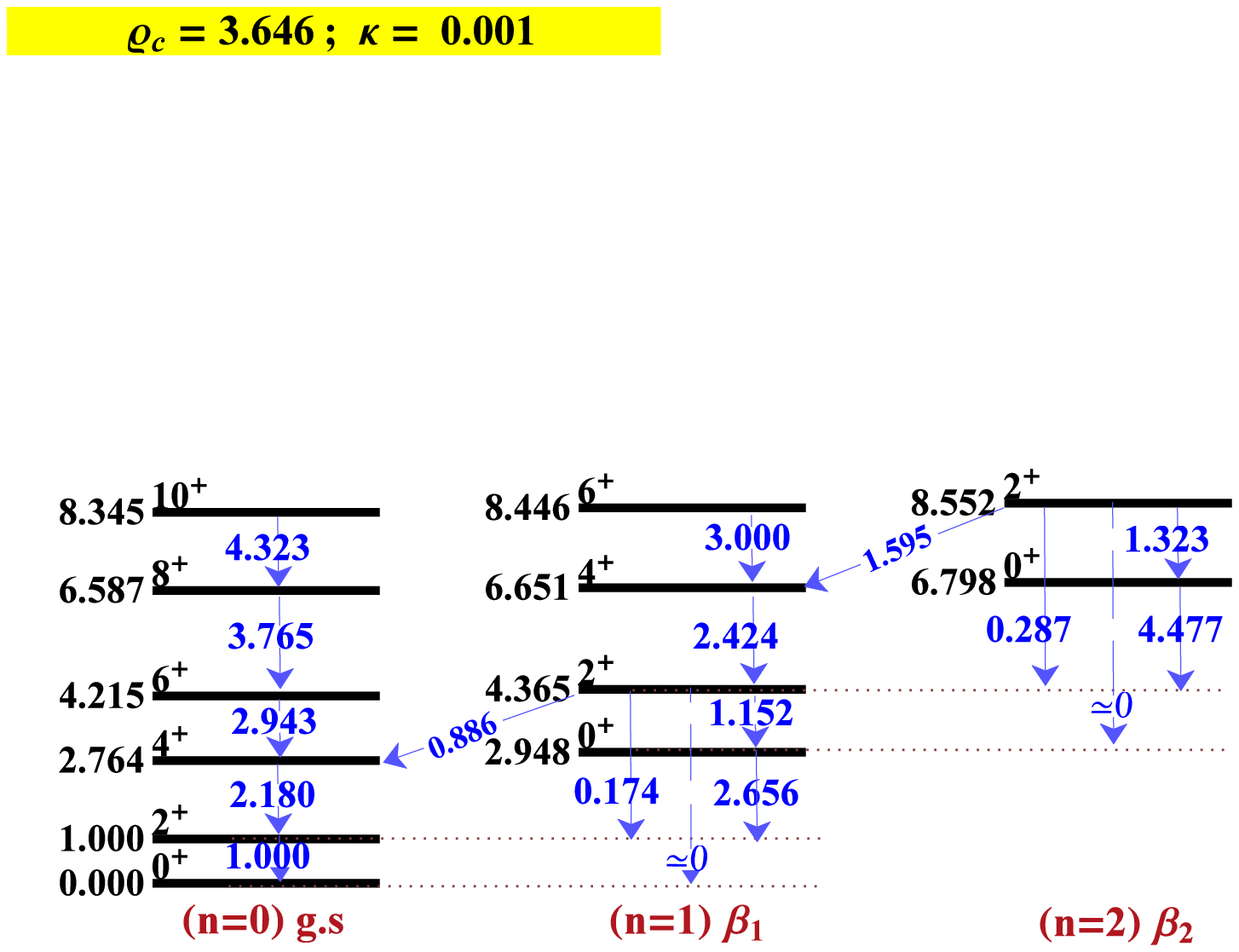}}{}\\
	{\includegraphics[scale=0.52]{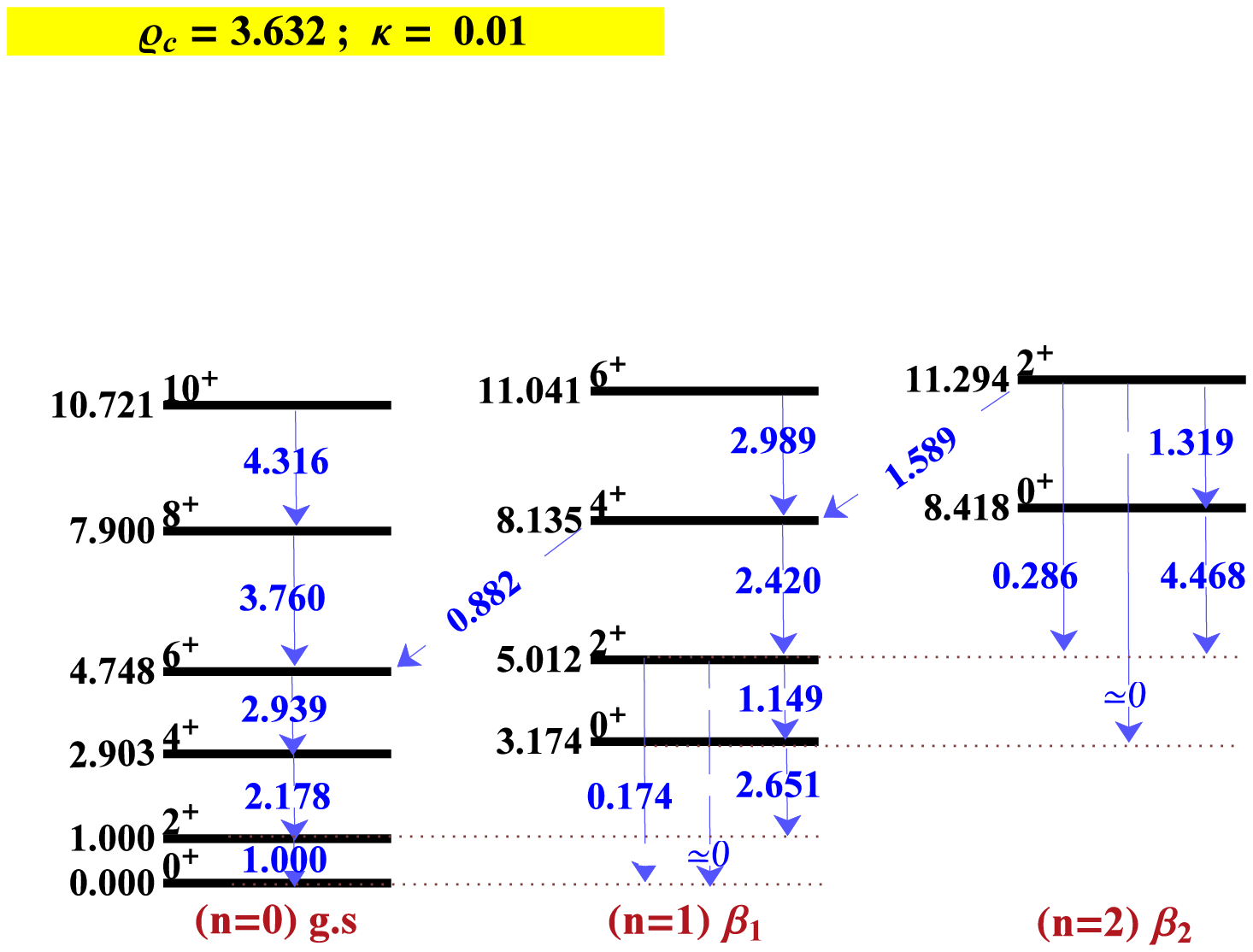}}{}{\includegraphics[scale=0.52]{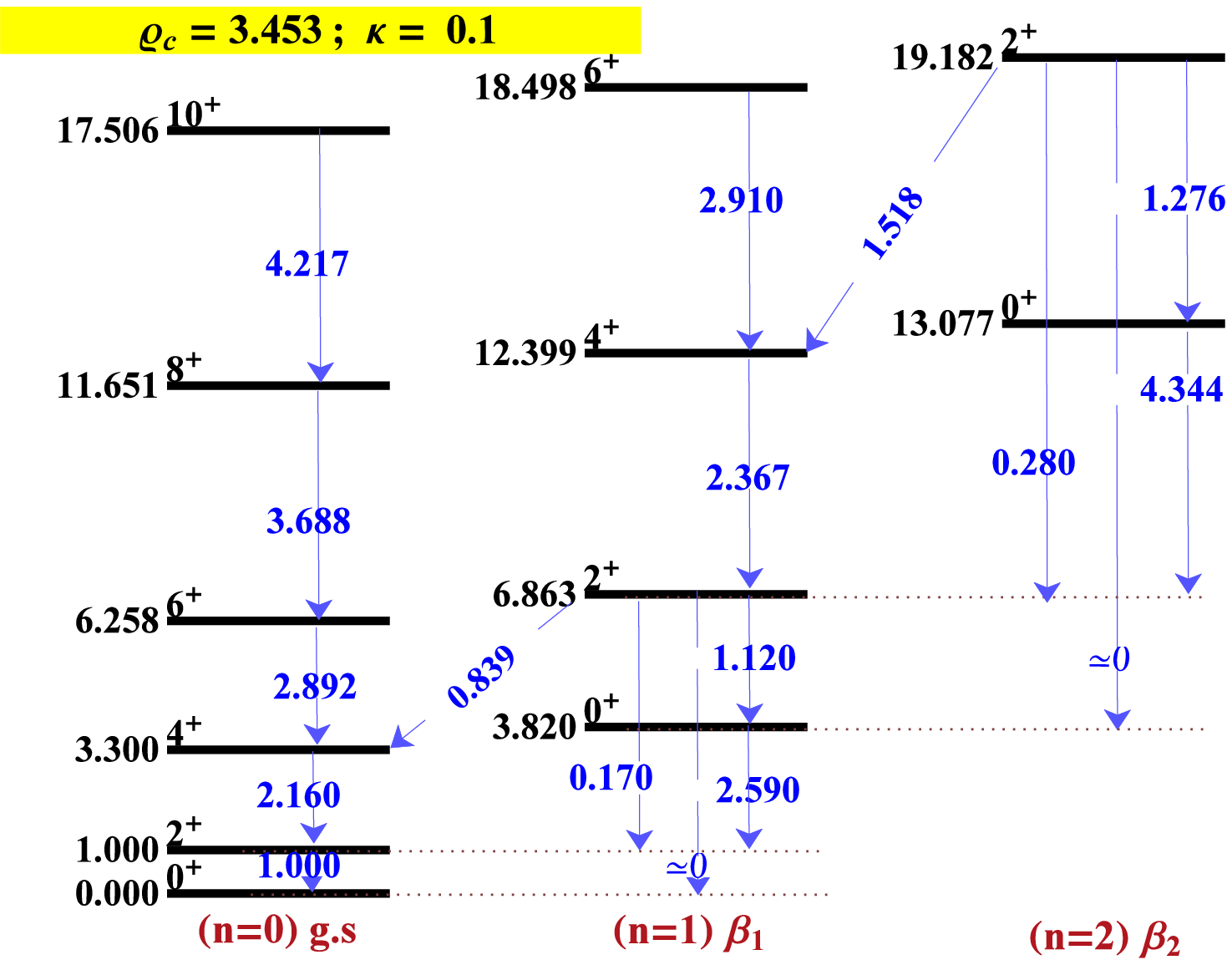}}{}\\
	\makeatother 
	\caption{{Energy spectra \eref{e27} and B(E2) transitions \eref{eq39}  for  $ (\varrho_{cs},\kappa) = (3.647 , 0.00) $, $ (3.646 , 0.001) $, $ (3.632 , 0.01) $, and $ (3.453 , 0.1) $  .}}
	\label{level}
\end{figure}
\egroup

The energy spectra \eref{e27} and some B(E2) transitions \eref{eq39} in the particular case of the critical point $\varrho_{c}$ are presented in \fref{level}  for   ($\varrho_{c} $,$ \kappa$) = (3.647 , 0.00) ,  (3.646 , 0.001) ,  (3.632 , 0.01) , and  (3.453 , 0.1) . If one analyzes now this figure, one can observe that the critical point varies slightly towards the lower values of $ \varrho $ as a function of the ML parameter $\kappa$. as can be deduced from table \ref{cpr}, for  ($ \kappa$ $ > 0.03 $), the critical point $\varrho_{c}$  $ \notin \left[ 2\sqrt{c^{(2)}_{0}}, 2\sqrt{c^{(2)}_{2}}\right]  $, this value   shows the limitation of the perturbation term.
\begin{table}[ht!]
	\caption{ The critical point $\varrho_{c}$ and the corresponding energy spectra R$_{0,4}$ are given as a function of the  minimal length parameter  $\kappa  $, the critical point region is defined by $  \left[ 2\sqrt{c^{(2)}_{0}}, 2\sqrt{c^{(2)}_{2}}\right] =\left[ 3.606, 3.872\right]  $ .}
	\vspace{0.3cm}
	\begin{center}
		\footnotesize
		\begin{tabular}{p{2.5cm}*{6}{@{\hskip.8mm}c@{\hskip.8mm}}}
			\br 
			$\kappa$  & 0 & 0.001&0.01&0.02&0.03&0.04 \\
			\mr
			$\varrho_{c}$  & 3.647 & 3.646&3.63&3.613&3.593&3.572 \\
				R$_{0,4}$  & 2.744 & 2.764&2.904&3.007&3.0804&3.135 \\
			\br	
		\end{tabular}
	\end{center}
	\label{cpr}
\end{table}
\normalsize
Moreover, as was already mentioned in \cite{b29}   in the case of  $\kappa =0$ ( without ML ), an approximate degeneracy appears for states of different angular momenta  $ (\Delta L = 4) $ belonging to different bands. Therefore,  the ML eliminate partially this approximate degeneracy.
Another important remark is that the states are grouped two by two in each band. However, 
 such a spectral behavior is not present in the results of the numerical diagonalization of a normal sextic potential, that is why it is necessary to consider the higher quasi-exact solvability order $  k>2 $ and verifying if this spectral characteristics persist in the critical point. This aspect will be analyzed in more details in a next work.
In what concerns B(E2) transition rates, both interband and intraband  transitions are decreasing when $\kappa$ is raised.

\subsection{\label{sec:level2}Energy Spectrum}
The theoretical predictions for the energy spectra of the g.s. band and the first two $ \beta $ bands are given by  \eref{e27} by fitting the model parameters on the experimental data using the quality measure : 
\begin{equation}
\sigma=\sqrt{\frac{\sum_{i=1}^{N^{\prime}}(E^{i}_{n,L}(exp)-E^{Corr,i}_{n,L}(th))^{2}}{N^{\prime} (E_{0,2})^{2}}},
\end{equation}
where $  N^{\prime} $ denotes the number of the states, while $E^{i}_{n,L}(exp)$ and $E^{Corr,i}_{n,L}(th)$ represent the experimental and theoretical energies of the $ i^{th} $ level, respectively. $E^{Corr}_{0,2}$ is the energy of the first excited level of the g.s. band.

The X(3)-SML model is applied to describe the available experimental data for the g.s. and the first two $ \beta $ bands of 35 nuclei,  namely, $^{98-108}$Ru, $^{100-102}$Mo, $^{116-130}$Xe,  $^{180-196}$Pt, $^{172}$Os, $^{146-150}$Nd, $^{132-134}$Ce, $^{154}$Gd, $^{156}$Dy and $^{150-152}$Sm.
The comparison for the energy spectra is shown in Tables \ref{tab1}, \ref{tab2}, \ref{tab3} and \ref{tab4}.
All these data are fitted involving the two free parameters $\varrho$ and $\kappa$. Their values and the corresponding rms being indicated in the same tables. The agreement with experiment, judged by the rms values, is very good for 83\% of  all studied nuclei. Indeed, the smallest rms (0.04) is found for the nuclei $ ^{98}$Ru. Moreover, the nuclei $^{122}$Xe, $^{100-106}$Ru, $^{134}$Ce and $^{184,188,190,194}$Pt have the rms values near to $0.20$, further, the rms values of the nuclei $^{118-120}$Xe, $^{102}$Mo, $^{108}$Ru, $^{180,192,196}$Pt and $^{150}$Sm are near 0.30. Also, those of $^{130}$Xe, $^{100}$Mo, $^{186}$Pt and $^{148}$Nd are near 0.40. Besides, those of $^{124,126}$Xe, $^{132}$Ce and $^{146}$Nd are near 0.50. Finally, those of $ ^{116}$Xe and $^{172}$Os are near 0.70. On the other hand, our numerical results show a good agreement and a better performance in respect to the sextic potential in the absence of ML case in the three bands: g.s. and two $\beta $ bands.

Concerning the nuclei $^{150}$Nd, $^{152}$Sm, $^{154}$Gd and $^{156}$Dy, which are known from the literature as good candidate for X(5), two different situations are presented here: the first one is the case of $\kappa =0 $, were the g.s. band prefers a $\gamma$-stable structure, while the situation of the first two $\beta$ bands is changed in a clear favor for $\gamma$-rigidity.
The second case corresponds to  $\kappa \ne 0 $, where experimental data are well reproduced only for the g.s. band. From  Table \ref{tab3}, we can see that the ML decreases the rms of those nuclei. As a matter of fact, the corresponding rms value of $^{150}$Nd is 0.80 instead of 1.10 in the case of $ \kappa=0 $.
Another interesting aspect of the present model is that the experimental data of some nuclei are well reproduced in the g.s. band more than in the $\beta$ band and conversely for the others. Indeed, for the nuclei whose $\varrho> 2\sqrt{c^{(2)}_{2}}$, the agreement between theory and experiment in the g.s. band is excellent, such as the $^{100}$Mo nucleus. However, in the case of the nuclei whose $\varrho < 2\sqrt{c^{(2)}_{0}}$, the $ \beta$ band being well described.
For the nuclei $^{128-130}$Xe and $^{192}$Pt,  the fitted values of $ \kappa $ vanish. In other words, they are not affected by the ML.
\begin{table}[ht!]
	\caption{The energy spectra for the ground band and the first two  $ \beta $  bands given by \eref{Rnl} are compared with the available experimental data  \cite{b44,b45,b46,b47,b48,b49,b51} of the nuclei $^{116-130}$Xe and with results available in Refs. \cite{b19,b29}
	In the first line of each nucleus are given the experimental data, in the second line are the corresponding $ X(3) $ \cite{b19}, in the third line are the corresponding X(3)-S \cite{b29}, while in the fourth line are the corresponding X(3)-SML. In brackets are indicated possible candidate energy states for the corresponding predicted data, which were not included in the fit.}
	\vspace{0.3cm}
	\begin{center}
		\footnotesize
		\begin{tabular}{p{3.3cm}*{13}{@{\hskip.6mm}c@{\hskip.6mm}}}
			\br  
			Nucleus \qquad&  R$_{0,4}$  &  R$_{0,6}$& R$_{0,8}$& R$_{0,10}$& R$_{1,0}$& R$_{1,2}$& R$_{1,4}$& R$_{1,6}$& R$_{2,0}$& R$_{2,2}$&$\sigma$&  $\varrho$ &  $\kappa$\\
			\mr
			{$ ^{116}$Xe } \quad{Exp}&2.33& 3.90& 5.62& 7.45& & 2.58& 3.96& 5.38& & & &&\\	
			\quad\qquad{X(3)}&2.44& 4.23& 6.35& 8.78& 2.87& 4.83& 7.37& 10.29& 7.65& 10.56& {	{2.48}	}& -&  0.0  \\
			\quad\qquad{X(3)-S}&2.28& 3.48& 4.97& 6.26& 2.25& 3.35& 4.76& 6.02& 4.75& 5.93& {	{0.72}	}& 6.62&  0.0  \\
			\quad\qquad{X(3)-SML}& 2.20& 3.49& 4.95& 6.47& 2.16& 3.33& 4.69& 6.15& 4.69& 6.05 & {	{0.68}	}& 18.30&  0.0016  \\
			\mr
			{$^{118}$Xe} \quad{Exp}&2.40& 4.14& 6.15& 8.35& 2.46& 3.64& 5.13& & (5.10)& &  &&\\
			\quad\qquad{X(3)}&2.44& 4.23& 6.35& 8.78& 2.87& 4.83& 7.37& 10.29& 7.65& 10.56& {	{0.99}	}& -&  0.0  \\	
			\quad\qquad{X(3)-S}&2.61& 3.94& 5.98& 7.48& 2.71& 3.98& 5.93& 7.44& 6.01& 7.47& {	{0.49}	}& 4.16&  0.0  \\
			\quad\qquad{X(3)-SML}& 2.35& 3.95& 5.91& 8.10& 2.41& 3.88& 5.74& 7.84& 5.84&7.83& {	{0.29}	}&14.96&  0.0070  \\
			\mr
			{$^{120}$Xe} \quad{Exp}&2.47& 4.33& 6.51& 8.90& 2.82& 3.95& 5.31& &(6.93)& & & &\\
			\quad\qquad{X(3)}&2.44& 4.23& 6.35& 8.78& 2.87& 4.83& 7.37& 10.29& 7.65& 10.56& {	{0.99}	}& -&  0.0  \\																		
			\quad\qquad{X(3)-S}&2.73& 4.11& 6.32& 7.91& 2.88& 4.21& 6.34& 7.94& 6.45& 8.01& {	{0.57}	}& 3.79&  0.0  \\
			\quad\qquad{X(3)-SML}& 2.40& 4.10& 6.23& 8.65& 2.50& 4.07& 6.09& 8.40& 6.22& 8.42& {	{0.36}	}& 14.89&  0.0097  \\
			\mr
			{$^{122}$Xe} \quad{Exp}&2.50& 4.43& 6.69& 9.18& 3.47& 4.51& & & (7.63)&  & &&\\
			\quad\qquad{X(3)}&2.44& 4.23& 6.35& 8.78& 2.87& 4.83& 7.37& 10.29& 7.65& 10.56& {	{0.36}	}& -&  0.0  \\																	
			\quad\qquad{X(3)-S}&2.63& 4.26& 6.74& 8.73& 3.29& 5.02& 7.33& 9.44& 7.72& 9.89& {	{0.30}	}& 2.11&  0.0  \\
			\quad\qquad{X(3)-SML}& 2.72& 4.33& 6.92& 9.02& 3.13& 4.78& 7.27& 9.46& 7.56& 9.76& {	{0.23}	}&2.71&  0.0020  \\
			\mr
			{$^{124}$Xe} \quad{Exp}&2.48& 4.37& 6.58& 8.96& 3.58& 4.60& 5.69& &(6.70) & (7.63) & &  & \\
			\quad\qquad{X(3)}&2.44& 4.23& 6.35& 8.78& 2.87& 4.83& 7.37& 10.29& 7.65& 10.56& {	{0.71}	}& -&  0.0  \\																				
			\quad\qquad{X(3)-S}&2.70& 4.17& 6.50& 8.23& 3.01& 4.48& 6.70& 8.50& 6.91& 8.69& {	{0.53}	}& 2.99&  0.0  \\
			\quad\qquad{X(3)-SML}& 2.44& 4.22& 6.51& 9.04& 2.57& 4.22& 6.39& 8.54& 6.54& 8.15& {	{0.49}	}& 13.03&  0.0132  \\
			\mr
			{$^{126}$Xe} \quad{Exp}&2.42& 4.21& 6.27& 8.64& 3.38& 4.32& 5.25& &(6.57)& & & &  \\
			\quad\qquad{X(3)}&2.44& 4.23& 6.35& 8.78& 2.87& 4.83& 7.37& 10.29& 7.65& 10.56& {	{0.85}	}& -& 0.0  \\																				
			\quad\qquad{X(3)-S}&2.72& 4.09& 6.29& 7.86& 2.86& 4.18& 6.30& 7.89& 6.40& 7.95& {	{0.55}	}& 3.83&  0.0  \\
			\quad\qquad{X(3)-SML}& 2.38& 4.06& 6.15& 8.51& 2.48& 4.02& 6.00& 8.26& 6.12& 8.27& {	{0.47}	}& 14.92&  0.0090  \\
			\mr
			{$^{128}$Xe} \quad{Exp}&2.33& 3.92& 5.67&& 7.60& 3.57& 4.52&&&(5.87) &  & &\\
			\quad\qquad{X(3)}&2.44& 4.23& 6.35& 8.78& 2.87& 4.83& 7.37& 10.29& 7.65& 10.56& {	{0.77}	}& -&  0.0  \\																					
			\quad\qquad{X(3)-S}&2.66& 4.01& 6.12& 7.65& 2.78& 4.07& 6.10& 7.65& 6.19& 7.69& {	{0.44}	}& 4.00&  0.0  \\
			\quad\qquad{X(3)-SML}&2.66& 4.01& 6.12& 7.65& 2.78& 4.07& 6.10& 7.65& 6.19& 7.69& {	{0.44}	}& 4.00&  0.0  \\	
			\mr
			{$^{130}$Xe} \quad{Exp}&2.25& 3.63& 5.03&& 3.35& (4.01)& (4.53)&&& &  & &\\
			\quad\qquad{X(3)}&2.44& 4.23& 6.35& 8.78& 2.87& 4.83& 7.37& 10.29& 7.65& 10.56& {	{0.77}	}& -&  0.0  \\																					
			\quad\qquad{X(3)-S}&2.41& 3.67& 5.37& 6.75& 2.43& 3.60& 5.23& 6.59& 5.26& 6.55& {	{0.50}	}& 5.18&  0.0  \\
			\quad\qquad{X(3)-SML}&2.41& 3.67& 5.37& 6.75& 2.43& 3.60& 5.23& 6.59& 5.26& 6.55& {	{0.50}	}& 5.18&  0.0  \\
				
			\br
		\end{tabular}
	\end{center}
	\label{tab1}
\end{table}
\normalsize
\begin{table}[ht!]
	\caption{The same as in \Tref{tab1}, but for the available experimental data \cite{b38,b39,b40,b41,b42,b43} of the nuclei $^{100-102}$Mo, $^{98-108}$Ru and $^{132}$Ce.}
	\begin{center}
		\footnotesize
		\begin{tabular}{p{3.3cm}*{13}{@{\hskip.6mm}c@{\hskip.6mm}}}
			\br  
			Nucleus \qquad&  R$_{0,4}$  &  R$_{0,6}$& R$_{0,8}$& R$_{0,10}$& R$_{1,0}$& R$_{1,2}$& R$_{1,4}$& R$_{1,6}$& R$_{2,0}$& R$_{2,2}$&$\sigma$&  $\varrho$ &  $\kappa$\\
			\mr
			{$^{100}$Mo} \quad{Exp}&2.12& 3.45& 4.91& 6.29& 1.30& 2.73& & & &&&&\\	
			\quad\qquad{X(3)}&2.44& 4.23& 6.35& 8.78& 2.87& 4.83& 7.37& 10.29& 7.65& 10.56& {	{1.63}	}& - & 0.0 \\														
			\quad\qquad{X(3)-S}&2.20& 3.37& 4.72& 5.95& 2.14& 3.20& 4.47& 5.66& 4.44& 5.54 & {	{0.43}	}& 8.63&  0.0  \\
			\quad\qquad{X(3)-SML }& 2.15& 3.40& 4.73& 6.12& 2.10& 3.21& 4.45& 5.79& 4.43& 5.66 & {	{0.39}	}& 38.20&  0.0005  \\
			\mr
			{$^{102}$Mo} \quad{Exp}&2.51& 4.48& 6.81& 9.41& 2.35& 3.86& & & &&&&\\
			\quad\qquad{X(3)}&2.44& 4.23& 6.35& 8.78& 2.87& 4.83& 7.37& 10.29& 7.65& 10.56& {	{0.56}	}& -&  0.0 \\
			\quad\qquad{X(3)-S}&2.67& 4.19& 6.57& 8.39& 3.09& 4.65& 6.90& 8.80& 7.16& 9.07& {	{0.63}	}& 2.65&  0.0  \\
			\quad\qquad{X(3)-SML}& 2.59& 4.28& 6.80& 9.27& 2.72& 4.34& 6.82& 9.30& 6.99& 9.40 & {	{0.27}	}& 5.32&  0.0104  \\
			\mr
			{$^{98}$Ru} \quad{Exp}&2.14& 3.41&4.79& 6.13& 2.03 & & & &&&&&\\
			\quad\qquad{X(3)}&2.44& 4.23& 6.35& 8.78& 2.87& 4.83& 7.37& 10.29& 7.65& 10.56& {	{1.48}	}& -&  0.0 \\
			\quad\qquad{X(3)-S}&2.21& 3.38& 4.73&5.97&2.15& 3.21& 4.49& 5.68& 4.46&5.57& {	{0.1}	}& 8.44&  0.0  \\
			\quad\qquad{X(3)-SML}& 2.17& 3.40& 4.76& 6.14& 2.12& 3.23& 4.50& 5.82& 4.47& 5.70& {	{0.04}	}& 16.00&  0.0010  \\
			\mr
			{$^{100}$Ru} \quad{Exp}& 2.27& 3.85& 5.67& 7.85& 2.10 & & & &&&&&\\
			\quad\qquad{X(3)}&2.44& 4.23& 6.35& 8.78& 2.87& 4.83& 7.37& 10.29& 7.65& 10.56& {	{0.65}	}& -&  0.0 \\
			\quad\qquad{X(3)-S}&2.58& 3.90& 5.88& 7.36& 2.66& 3.92& 5.82& 7.31& 5.89& 7.32& {	{0.37}	}& 4.28&  0.0  \\
			\quad\qquad{X(3)-SML}& 2.33& 3.83& 5.70& 7.70& 2.37& 3.75& 5.53& 7.44& 5.60& 7.39& {	{0.14}	}& 10.95&  0.006  \\
			\mr
			{$^{102}$Ru} \quad{Exp}&2.33& 3.94& 5.70& 7.23& 1.99& & & &&&&&\\
			\quad\qquad{X(3)}&2.44& 4.23& 6.35& 8.78& 2.87& 4.83& 7.37& 10.29& 7.65& 10.56& {	{0.86}	}& -&  0.0 \\
			
			\quad\qquad{X(3)-S}&2.50& 3.79& 5.64& 7.08& 2.55& 3.77& 5.54& 6.98& 5.59& 6.96& {	{0.28}	}&4.63&  0.0  \\
			\quad\qquad{X(3)-SML}& 2.32& 3.83& 5.66& 7.23& 2.36& 3.74& 5.39& 5.06& 5.39& 2.26& {	{0.17}	}& 12.68& 0.0057  \\
			\mr
			{$^{104}$Ru} \quad{Exp}&2.48& 4.35& 6.48& 8.69& (2.76)& 4.23& 5.81 & & & &&&\\	
			\quad\qquad{X(3)}&2.44& 4.23& 6.35& 8.78& 2.87& 4.83& 7.37& 10.29& 7.65& 10.56& {	{0.69}	}& -&  0.0 \\											
			\quad\qquad{X(3)-S}&2.74& 4.14& 6.41& 8.02& 2.92& 4.28& 6.45& 8.10& 6.58& 8.19& {	{0.40}	}& 3.62&  0.0  \\
			\quad\qquad{X(3)-SML}& 2.53& 4.10& 6.37& 8.54& 2.62& 4.11& 6.33& 8.50& 6.47& 8.56& {	{0.25}	}& 5.42&  0.0071  \\
			\mr
			{$^{106}$Ru} \quad{Exp}&2.66& 4.80& 7.31& 10.02& 3.67& & & &&&&&\\
			\quad\qquad{X(3)}&2.44& 4.23& 6.35& 8.78& 2.87& 4.83& 7.37& 10.29& 7.65& 10.56& {{0.83}	}& -&  0.0 \\																\quad\qquad{X(3)-S}&2.55& 4.43& 7.14& 9.55& 4.10& 6.20& 8.55& 11.19& 9.36& 12.20 & {	{0.34}	}& 1.16&  0.0  \\
			\quad\qquad{X(3)-SML}& 2.65& 4.53& 7.42&10.01& 3.67& 5.73& 8.43& 11.23& 9.03& 11.97& {	{0.13}	}& 1.61&  0.0027  \\
			\mr
			{$^{108}$Ru} \quad{Exp}&2.75& 5.12& 8.02& 11.31& 4.03& & & &&&&&\\
			\quad\qquad{X(3)}&2.44& 4.23& 6.35& 8.78& 2.87& 4.83& 7.37& 10.29& 7.65& 10.56& {{1.51}	}& -&  0.0 \\														
			\quad\qquad{X(3)-S}&2.49& 4.50& 7.30& 9.92& 4.75& 7.01& 9.31& 12.22& 10.43& 13.65 & {	{0.83}	}& 0.73&  0.0  \\
				\quad\qquad{X(3)-SML}& 2.67& 4.79& 8.11& 11.32& 4.01&6.37& 9.50& 13.02& 10.29& 14.0&  {	{0.16}	}& 1.27&  0.0061  \\
					\mr
					{$^{132}$Ce} \quad{Exp}&2.64& 4.74& 7.16& 9.71& 3.56& 4.60& 5.94 & & & &&&\\
					\quad\qquad{X(3)}&2.44& 4.23& 6.35& 8.78& 2.87& 4.83& 7.37& 10.29& 7.65& 10.56& {{0.79}	}& -&  0.0 \\
					\quad\qquad{X(3)-S}&2.65& 4.23& 6.65& 8.56& 3.18& 4.83& 7.11& 9.11& 7.43& 9.47& {	{0.70}	}& 2.36&  0.0  \\		
					\quad\qquad{X(3)-SML}& 2.49& 4.40& 6.85& 9.71& 2.66& 4.43& 6.75& 9.48& 6.95& 9.56& {	{0.50}	}& 16.11& 0.0173  \\
			\br
		\end{tabular}
	\end{center}
	\label{tab2}
\end{table}
\normalsize
\begin{table}[ht!]
	\caption{The same as in \Tref{tab1}, but for the available experimental data \cite{b52,b53,b60,b61,b62,b63,b64,b65,b66} of the nuclei $^{134}$Ce, $^{172}$Os and $^{180-192}$Pt.}
	\vspace{0.3cm}
	\begin{center}
		\footnotesize
		\begin{tabular}{p{3.3cm}*{13}{@{\hskip.6mm}c@{\hskip.6mm}}}
			\br  
			Nucleus \qquad&  R$_{0,4}$  &  R$_{0,6}$& R$_{0,8}$& R$_{0,10}$& R$_{1,0}$& R$_{1,2}$& R$_{1,4}$& R$_{1,6}$& R$_{2,0}$& R$_{2,2}$&$\sigma$&  $\varrho$ &  $\kappa$\\
		
			\mr
			{$^{134}$Ce} \quad{Exp}&2.56& 4.55& 6.87& 9.09& 3.75& 4.80& & & & &&&\\
			\quad\qquad{X(3)}&2.44& 4.23& 6.35& 8.78& 2.87& 4.83& 7.37& 10.29& 7.65& 10.56& {{0.46}	}& -&  0.0 \\
			\quad\qquad{X(3)-S}&2.61& 4.30& 6.82& 8.89& 3.42& 5.23& 7.56& 9.77& 8.01& 10.31& {	{0.26}	}& 1.89&  0.0  \\
			\quad\qquad{X(3)-SML}& 2.65& 4.33& 6.92& 9.06&3.32& 5.10& 7.54& 9.82& 7.95& 10.29& {	{0.24}	}&2.10& 0.0010  \\
			\mr
			{$^{172}$Os} \quad{Exp}&2.66& 4.63& 6.70& 8.89& 3.33& 3.56& 5.00& 6.81&&&&&\\
			\quad\qquad{X(3)}&2.44& 4.23& 6.35& 8.78& 2.87& 4.83& 7.37& 10.29& 7.65& 10.56& {{1.58}	}& -&  0.0 \\									
			\quad\qquad{X(3)-S}&2.65& 4.00& 6.10& 7.63& 2.77& 4.06& 6.07& 7.62& 6.16& 7.66& {	{0.77}	}& 4.02&  0.0  \\
			\quad\qquad{X(3)-SML}& 2.35& 3.91& 5.86& 7.91& 2.41& 3.84& 5.69& 7.37& 5.78& 6.90 & {	{0.70}	}& 11.17& 0.0072  \\
			\mr
			{$^{180}$Pt} \quad{Exp}&2.68& 4.94& 7.71& 10.93& 3.12& 5.62& 8.15& 10.77& (7.69)& & &  &\\
			\quad\qquad{X(3)}&2.44& 4.23& 6.35& 8.78& 2.87& 4.83& 7.37& 10.29& 7.65& 10.56& {1.03}	& -&  0.0 \\									
			\quad\qquad{X(3)-S}&2.56& 4.40& 7.07& 9.41& 3.92& 5.96& 8.32& 10.86& 9.03& 11.75& {	{0.69}	}& 1.31&  0.0  \\
			\quad\qquad{X(3)-SML}& 2.86& 4.70& 7.84& 10.63& 3.22& 5.10& 8.20& 11.11& 8.52& 11.43& {	{0.27}	}& 3.05& 0.0085  \\
			\mr   
			{$^{182}$Pt} \quad{Exp}&2.71& 5.00& 7.78& 10.96& 3.22& 5.53& 8.00& 10.64& (7.43)& & &  &\\
			\quad\qquad{X(3)}&2.44& 4.23& 6.35& 8.78& 2.87& 4.83& 7.37& 10.29& 7.65& 10.56& {{1.04}	}& -&  0.0 \\									
			\quad\qquad{X(3)-S}&2.57& 4.40& 7.05& 9.37& 3.88& 5.90& 8.26& 10.77& 8.95& 11.64& {	{0.72}	}& 1.35& 0.0  \\
			\quad\qquad{X(3)-SML}& 2.90& 4.74& 7.87& 10.67& 3.17&4.99& 8.10& 10.98& 8.37& 11.23 & {	{0.28}	}& 3.67& 0.0099  \\
			\mr
			{$^{184}$Pt} \quad{Exp}&2.67& 4.90& 7.55& 10.47& 3.02& 5.18& 7.57& 11.04&& &  & &\\
			\quad\qquad{X(3)}&2.44& 4.23& 6.35& 8.78& 2.87& 4.83& 7.37& 10.29& 7.65& 10.56& {	{0.83}	}& -&  0.0 \\									
			\quad\qquad{X(3)-S}&2.57& 4.38& 7.01& 9.29& 3.79& 5.77& 8.13& 10.59& 8.77& 11.39& {	{0.65}	}& 1.44&  0.0  \\
			\quad\qquad{X(3)-SML}& 2.85& 4.66& 7.72& 10.41& 3.18& 5.01& 8.02& 10.82& 8.33& 11.11 & {	{0.23}	}& 3.18& 0.0078  \\
			\mr
			{$^{186}$Pt} \quad{Exp}&2.56& 4.58& 7.01& 9.70& 2.46& 4.17& 6.38& 8.36&& &  & &\\
			\quad\qquad{X(3)}&2.44& 4.23& 6.35& 8.78& 2.87& 4.83& 7.37& 10.29& 7.65& 10.56& {	{0.92}	}& -&  0.0 \\									
			\quad\qquad{X(3)-S}&2.68& 4.18& 6.53& 8.30& 3.04& 4.56& 6.80& 8.64& 7.03& 8.87& {	{0.62}	}& 2.28&  0.0  \\
			\quad\qquad{X(3)-SML}&2.56& 4.24& 6.72& 9.17& 2.69& 4.29& 6.71& 9.16& 6.88& 9.25 & {	{0.40}	}& 5.59& 0.0105  \\
			\mr
			{$^{188}$Pt} \quad{Exp}&2.53& 4.46& 6.71& 9.18& 3.01& 4.20&& & & & &&\\
			\quad\qquad{X(3)}&2.44& 4.23& 6.35& 8.78& 2.87& 4.83& 7.37& 10.29& 7.65& 10.56& {	{0.36}	}& -&  0.0 \\									
			\quad\qquad{X(3)-S}&2.65& 4.22& 6.64& 8.53& 3.17& 4.80& 7.08& 9.06& 7.39& 9.41& {	{0.38}	}& 2.40&  0.0  \\
			\quad\qquad{X(3)-SML}& 2.63& 4.28& 6.80& 9.16& 2.77& 4.36& 6.83& 9.21& 7.00& 9.32 & {	{0.15}	}& 4.76& 0.0082  \\		
			\mr
			{$^{190}$Pt} \quad{Exp}&2.49& 4.35& 6.47& 8.57& 3.11& 4.07&& &(5.65) & & &&\\
			\quad\qquad{X(3)}&2.44& 4.23& 6.35& 8.78& 2.87& 4.83& 7.37& 10.29& 7.65& 10.56& {	{0.34}	}& -&  0.0 \\									
			\quad\qquad{X(3)-S}&2.69& 4.17& 6.50& 8.24& 3.01& 4.50& 6.72& 8.52& 6.93& 8.72& {	{0.25}	}& 2.96&  0.0  \\
			\quad\qquad{X(3)-SML}& 2.55& 4.13& 6.46& 8.67& 2.65
			& 4.16& 6.43& 8.64& 6.57& 8.71& {	{0.22}	}& 5.28& 0.0073  \\	
				\mr
				{$^{192}$Pt} \quad{Exp}&2.48& 4.31& 6.38& 8.62& 3.78& 4.55&& & & & &&\\
				\quad\qquad{X(3)}&2.44& 4.23& 6.35& 8.78& 2.87& 4.83& 7.37& 10.29& 7.65& 10.56& {	{0.34}	}& -&  0.0 \\									
				\quad\qquad{X(3)-S}&2.65& 4.23& 6.66& 8.57& 3.19& 4.84& 7.13& 9.14& 7.45& 9.50& {	{0.30}	}& 2.34&  0.0  \\
				\quad\qquad{X(3)-SML}&2.65& 4.23& 6.66& 8.57& 3.19& 4.84& 7.13& 9.14& 7.45& 9.50& {	{0.30}	}& 2.34&  0.0  \\
			
			\br
		\end{tabular}
	\end{center}
	\label{tab3}
\end{table}
\normalsize
\begin{table}[ht!]
	\caption{The same as in \Tref{tab1}, but for the available experimental data \cite{b68,b69,b54,b55,b56,b57,b58,b59}	of the nuclei  $^{194-196}$Pt, $^{146-150}$Nd, $^{150-152}$Sm, $^{154}$Gd and $ ^{156}$Dy.}
	\vspace{0.3cm}
	\begin{center}
		\footnotesize
		\begin{tabular}{p{3.3cm}*{13}{@{\hskip.6mm}c@{\hskip.6mm}}}
			\br

			Nucleus \qquad&  R$_{0,4}$  &  R$_{0,6}$& R$_{0,8}$& R$_{0,10}$& R$_{1,0}$& R$_{1,2}$& R$_{1,4}$& R$_{1,6}$& R$_{2,0}$& R$_{2,2}$&$\sigma$&  $\varrho$ &  $\kappa$\\
		
			\mr
			{$^{194}$Pt} \quad{Exp}&2.47& 4.30& 6.39& 8.67& 3.23& 4.60& & & &&&&\\
			\quad\qquad{X(3)}&2.44& 4.23& 6.35& 8.78& 2.87& 4.83& 7.37& 10.29& 7.65& 10.56& {	{0.18}	}& -&  0.0 \\									
			\quad\qquad{X(3)-S}&2.64& 4.24& 6.69& 8.63& 3.86& 4.91& 7.20& 9.25& 7.55& 9.65& {	{0.19}	}& 2.25&  0.0  \\
			\quad\qquad{X(3)-SML} &2.67& 4.22& 6.64& 8.51& 3.13& 4.73& 7.01& 8.98& 7.30& 9.29& {	{0.16}	}& 2.53&  0.0002  \\
			\mr	
			{$^{196}$Pt} \quad{Exp}&2.47& 4.29& 6.33& 8.56& 3.19& 3.83& & & &&&&\\
			\quad\qquad{X(3)}&2.44& 4.23& 6.35& 8.78& 2.87& 4.83& 7.37& 10.29& 7.65& 10.56& {	{0.44}	}& -&  0.0 \\										
			\quad\qquad{X(3)-S}&2.72& 4.15& 6.45& 8.13& 2.96& 4.39& 6.59& 8.31& 6.76& 8.46& {	{0.33}	}& 3.25&  0.0  \\
			\quad\qquad{X(3)-SML }&2.61& 4.12& 6.43& 8.43& 2.73& 4.18& 6.42& 8.44& 6.55& 8.52& {	{0.26}	}& 4.53& 0.0043  \\
			\mr
			{$^{146}$Nd} \quad{Exp}& 2.30& 3.92& 5.72& 7.32& 2.02& 2.87& 3.85& & & & &&\\
			\quad\qquad{X(3)}&2.44& 4.23& 6.35& 8.78& 2.87& 4.83& 7.37& 10.29& 7.65& 10.56& {	{1.67}	}& -&  0.0 \\	
			\quad\qquad{X(3)-S}&2.34& 3.56& 5.14& 6.46& 2.32& 3.45& 4.96& 6.26& 4.96& 6.19& {	{0.64}	}& 5.88&  0.0  \\
			\quad\qquad{X(3)-SML}& 2.24& 3.58& 5.15& 6.76& 2.22& 3.44& 4.92& 6.38& 4.93& 6.17& {	{0.57}	}& 13.12& 0.0026  \\
			\mr
			{$^{148}$Nd} \quad{Exp}&2.49& 4.24& 6.15& 8.19& 3.04& 3.88& 5.32& 7.12& (5.30) &&&&\\
			\quad\qquad{X(3)}&2.44& 4.23& 6.35& 8.78& 2.87& 4.83& 7.37& 10.29& 7.65& 10.56& {	{1.40}	}& -&  0.0 \\	
			\quad\qquad{X(3)-S}&2.61& 3.95& 5.98& 7.49& 2.71& 3.98& 5.94& 7.46& 6.02& 7.48& {	{0.39}	}& 4.15&  0.0  \\
			\quad\qquad{X(3)-SML}& 2.52& 3.92& 5.97& 7.71& 2.59& 3.92& 5.90& 7.64& 5.99& 7.67& {	{0.38}	}& 4.82& 0.0025  \\
			\mr
			{$^{150}$Nd} \quad{Exp}&2.93& 5.53& 8.68& 12.28& 5.19& 6.53& 8.74& 11.83& (13.35)& (5.30) &&&\\
			\quad\qquad{X(3)}&2.44& 4.23& 6.35& 8.78& 2.87& 4.83& 7.37& 10.29& 7.65& 10.56& {	{2.00}	}& -&  0.0 \\	
			\quad\qquad{X(3)-S}&2.49& 4.50& 7.30& 9.92& 4.75& 7.01& 9.31& 12.22& 10.43& 13.65& {	{1.10}	}& 0.73&  0.0  \\
			\quad\qquad{X(3)-SML}& 2.73&4.82&8.19& 11.41& 3.78& 6.07& 9.33& 12.83& 10.00& 13.65 & {	{0.80}	}& 1.56& 0.0074  \\
			\mr
			{$^{150}$Sm} \quad{Exp}&2.32& 3.83& 5.50& 7.29& 2.22& 3.13& 4.34& 6.31& (3.76) &&&&\\
			\quad\qquad{X(3)}&2.44& 4.23& 6.35& 8.78& 2.87& 4.83& 7.37& 10.29& 7.65& 10.56& {	{1.98}	}& -&  0.0 \\	
			\quad\qquad{X(3)-S}&2.36& 3.60& 5.23& 6.57& 2.36& 3.51& 5.06& 6.39& 5.07& 6.33& {	{0.41}	}& 5.58&  0.0  \\
			\quad\qquad{X(3)-SML }& 2.23& 3.60& 5.16& 6.85& 2.22& 3.46& 4.93& 6.53& 4.95& 6.45 & {	{0.33}	}& 21.80& 0.0021  \\	
			\mr
			{$^{152}$Sm} \quad{Exp}&3.01& 5.80& 9.24& 13.21& 5.62& 6.65& 8.40& 10.76& 8.89 &10.62&&&\\
			\quad\qquad{X(3)}&2.44& 4.23& 6.35& 8.78& 2.87& 4.83& 7.37& 10.29& 7.65& 10.56& 1.62& -&  0.0 \\		
			\quad\qquad{X(3)-S}&2.55& 4.43& 7.14& 9.55& 4.10& 6.20& 8.55& 11.19& 9.36& 12.20& 1.59& 1.16&  0.0  \\
			\quad\qquad{X(3)-SML}& 2.95& 4.91& 8.31& 11.48& 3.24& 5.21& 8.59& 11.85& 8.91& 12.14 & 1.26& 3.70& 0.0287  \\		
			\mr
			{$^{154}$Gd} \quad{Exp}&3.01& 5.83& 9.30& 13.3& 5.53& 6.63& 8.51& 11.10& 9.60 &11.52&&&\\
			\quad\qquad{X(3)}&2.44& 4.23& 6.35& 8.78& 2.87& 4.83& 7.37& 10.29& 7.65& 10.56& 1.51& -&  0.0 \\		
			\quad\qquad{X(3)-S}&2.53& 4.46& 7.21& 9.70& 4.33& 6.50& 8.84& 11.58& 9.75& 12.74& 1.51& 0.99&  0.0  \\
			\quad\qquad{X(3)-SML }& 2.97& 5.01& 8.55& 11.90& 3.28& 5.33& 8.86& 12.32& 9.21& 12.64 & 1.14& 3.68& 0.0165  \\		
			\mr
			{$^{156}$Dy} \quad{Exp}&2.93& 5.59& 8.82& 12.52& 4.90& 6.01& 7.90& 10.43& 10.00 &&&&\\
			\quad\qquad{X(3)}&2.44& 4.23& 6.35& 8.78& 2.87& 4.83& 7.37& 10.29& 7.65& 10.56&1.28& -&  0.0 \\		
			\quad\qquad{X(3)-S}&2.53& 4.46& 7.20& 9.68& 4.31& 6.46& 8.81& 11.53& 9.71& 12.68& 1.28& 1.01&  0.0  \\
			\quad\qquad{X(3)-SML}& 2.96& 4.96& 8.41& 11.64& 3.26& 5.26& 8.71& 12.04& 9.04& 12.35 & 	0.99& 3.65& 0.0148  \\	
			\br
		\end{tabular}
	\end{center}
	\label{tab4}
\end{table}
\normalsize
\begin{table}[ht!]
	\caption{The roots mean square obtained with our model X(3)-SML are compared with the values taken from the models X(3)-S \cite{b29}, X(3)-D-ML  \cite{b32}, X(3)-H-ML\cite{b33} and X(3)-SML for the nuclei :$^{104}$Ru, $^{120-124}$Xe, $^{148}$Nd, $^{150}$Sm, $^{180-186}$Pt and $^{196}$Pt.}
	\begin{center}
		\footnotesize
		\begin{tabular}{p{4.5cm}*{4}{@{\hskip.8mm}c@{\hskip.8mm}}}
			\br
			Nucleus &$\sigma_{X(3)-H-ML}$\cite{b33} & $\sigma_{X(3)-D-ML}$ \cite{b32} &$\sigma_{X(3)-S}$ \cite{b29} &$\sigma_{X(3)-SML}$ \\
			\mr
			{ $^{104}$Ru}&0.63& $ - $ &0.40 &0.25 \\
			{ $^{120}$Xe}&0.70& $ - $ &0.57 &0.36  \\
			{ $^{122}$Xe}&0.58& $ - $ &0.30 &0.23  \\
			{ $^{124}$Xe}&0.47& $ - $ &0.53 &0.49  \\
			{ $^{148}$Nd}&0.49& $ - $ &0.39 &0.38 \\
			{ $^{150}$Sm}&0.64& $ - $ &0.41 &0.33  \\
			{ $^{180}$Pt}& $ - $ & 0.68 &0.69 &0.27  \\
			{ $^{182}$Pt}&$ - $& 0.47 &0.72 &0.28  \\
			{ $^{184}$Pt}&$ - $& 0.45 &0.65 &0.23 \\
			{ $^{186}$Pt}&$ - $& 0.47 &0.62 &0.37  \\
			{ $^{192}$Pt}&0.41& $ - $ &0.30 &0.30  \\
			{ $^{196}$Pt}&0.60& $ - $ &0.33 &0.26  \\
			\br
		\end{tabular}
	\end{center}
	\label{comp}
\end{table}	
\normalsize
Such facts are clearly illustrated in tables \ref{tab1}, \ref{tab2},   \ref{tab3} and  \ref{tab4}, where one can also observe that the ML effect is more apparent at higher angular momenta.  
It is important to compare our model with other solutions presented in Table \ref{tabmodels}, in order to evaluate the sextic potential in the presence of the ML. This comparison is presented in Table \ref{comp} and shows that in the presence of ML, the obtained results for rms with a sextic potential are fairly better than those obtained with both Hulthen and Davidson potentials.
 
\bgroup
\begin{figure}
	{\includegraphics[scale=0.85]{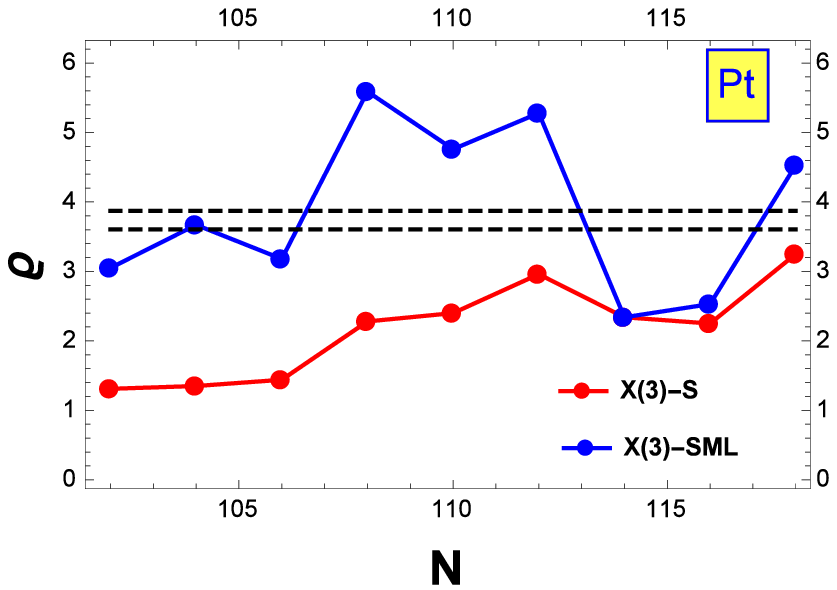}}{}	{\includegraphics[scale=0.85]{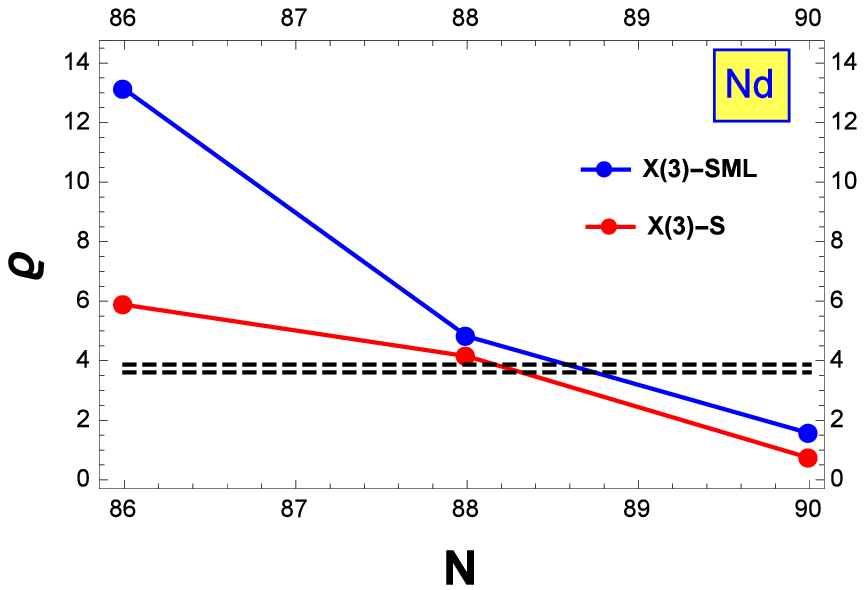}}{}\\
	{\includegraphics[scale=0.85]{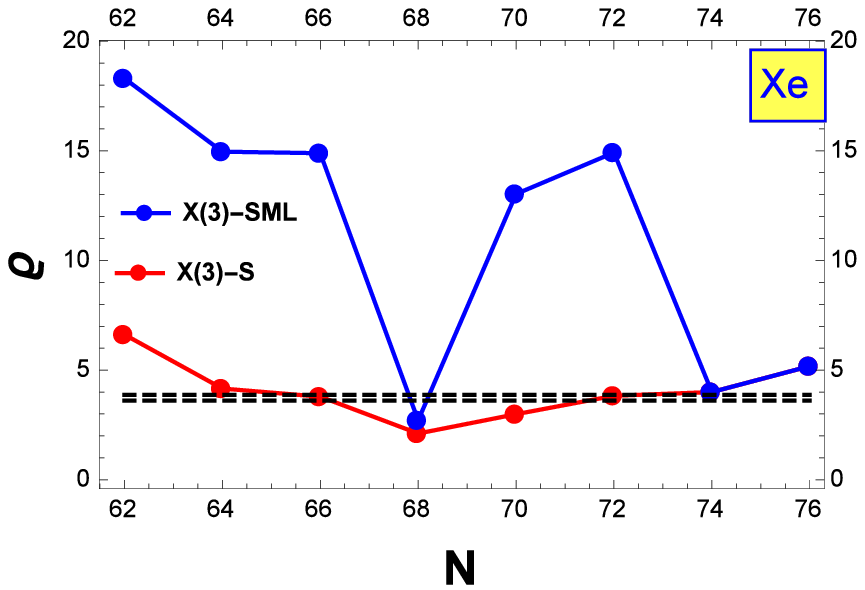}}{}	{\includegraphics[scale=0.85]{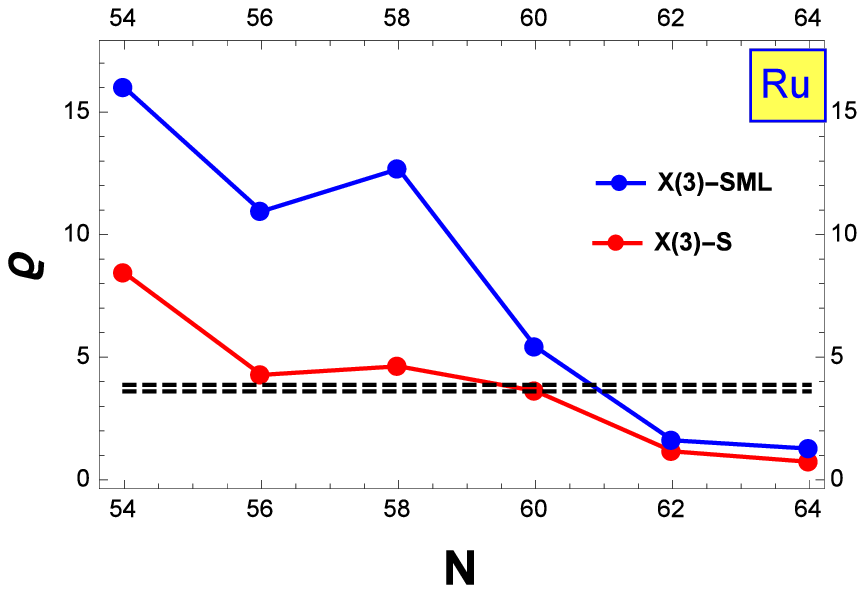}}{}
	\makeatother 
	\caption{{The free parameter $ \varrho $ as a function of the neutron number N for isotopes of Ru, Xe, Pt and Nd, where $ \kappa = 0 $ (Red lines) and $ \kappa \ne 0 $ (Blue lines). Continuous lines indicate the shape evolution within the isotope chain, while the dashed ones are for the critical point region.}}
	\label{Isotropic}
\end{figure}
\egroup

Moreover, it is interesting to investigate the ML effect on the shape phase transition. For this purpose, the fitted values $ \varrho $ are plotted in \fref{Isotropic} as a function of the neutron number $ N $ for the most isotopic chains considered in the present paper, namely: Ru, Xe, Nd and Pt. The area between the dashed lines corresponds to the critical point region, while the regions above and below the dashed lines correspond to the spherical and  deformed regions, respectively.
For the isotopes of Ru, a phase transition takes place from a spherical shape (lighter nuclei) to a deformed one (heaviest nuclei). In the absence of ML ($ \kappa=0 $), the horizontal dashed lines are crossed going from  $ ^{102}$Ru to $ ^{104}$Ru, while for ($ \kappa \ne 0 $) this is happening between $ ^{104}$Ru and $ ^{106}$Ru. Physically speaking, one can say that the phase transition takes place faster in the absence of ML than in its presence.
The phase transition for the isotopes of Nd seems to be similar to that for Ru crossing once the dashed lines and having the lightest isotopes and the heaviest ones situated above and below the lines.
In what concerns the isotopes of Xe, both for $ \kappa=0 $ and $ \kappa \ne 0 $, they cross twice the dashed lines. Indeed, two shape phase transitions converge to $ ^{122}$Xe. The first one is $^{116}Xe \rightarrow \ ^{122}Xe$ ( from the lightest isotopes towards the medium ones ), while the second one is $^{130}Xe \rightarrow \ ^{122}Xe$ (from the heaviest isotopes towards the medium ones) and $ ^{122}$Xe plays the role of critical point of a transition between the two transition's arms of Xe.
The major effects of the ML are observed for the Pt isotopes, where the shape evolution changed dramatically comparing to the description offered by the case where ML is absent. According to $ \kappa=0 $, there is no phase transition from spherical shape to a deformed one for the isotopes of Pt, which is not the case of $ \kappa \ne 0 $, for which one has three intersections of the critical point: once from the $ ^{186}$Pt towards  $ ^{184}$Pt and the second time from  $ ^{190}$Pt to  $ ^{192}$Pt and from  $ ^{196}$Pt to  $ ^{194}$Pt.

Other remarks can be drawn by analyzing the  \fref{Isotropic}, one can observe that the results concerning the shape of nuclei in the presence of ML are totally different from those in its absence.
The best candidates for the critical point of the $ \gamma$-rigid prolate harmonic vibrator to $ \gamma $-rigid anharmonic vibrator shape phase transition are found to be $ ^{182}$Pt, $ ^{152}$Sm, $ ^{154}$Gd, $ ^{156}$Dy, respectively. Other possible candidates can be also considered: $ ^{184}$Pt and $ ^{128}$Xe.

Note that a higher order $  k>2 $  could offer, in this way, a more realistic description of the isotopic chains and it will be the topic of our future work.
\bgroup
\begin{figure}
	{\includegraphics[scale=0.85]{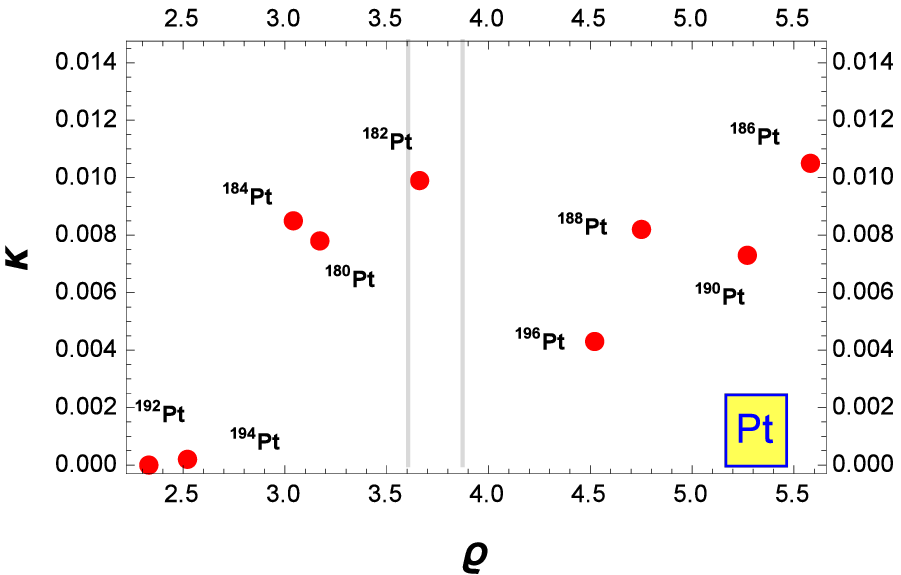}}{}	{\includegraphics[scale=0.85]{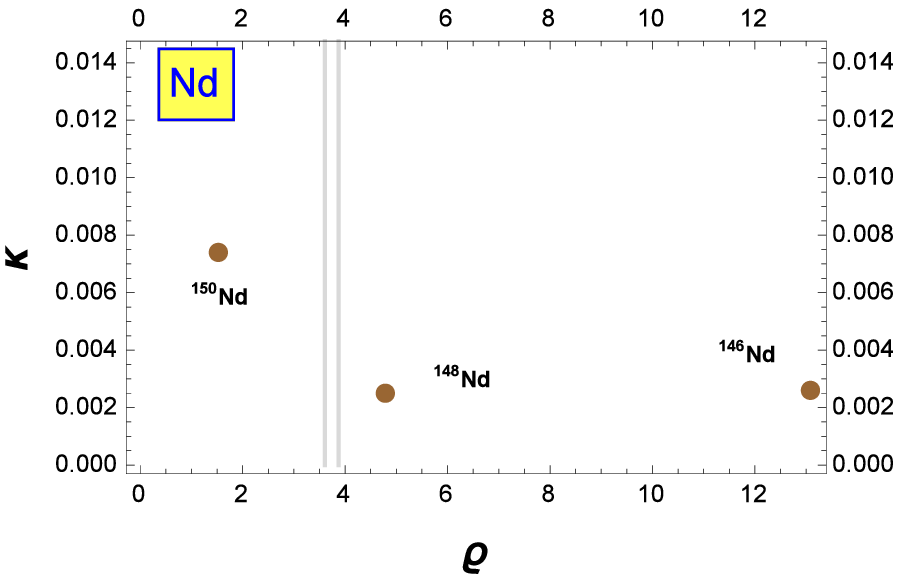}}{}\\
	{\includegraphics[scale=0.85]{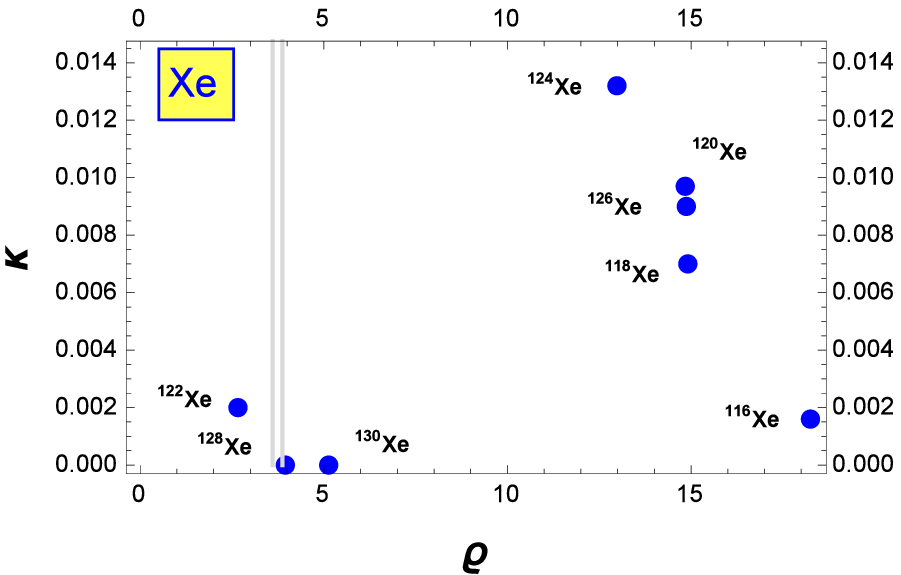}}{}	{\includegraphics[scale=0.85]{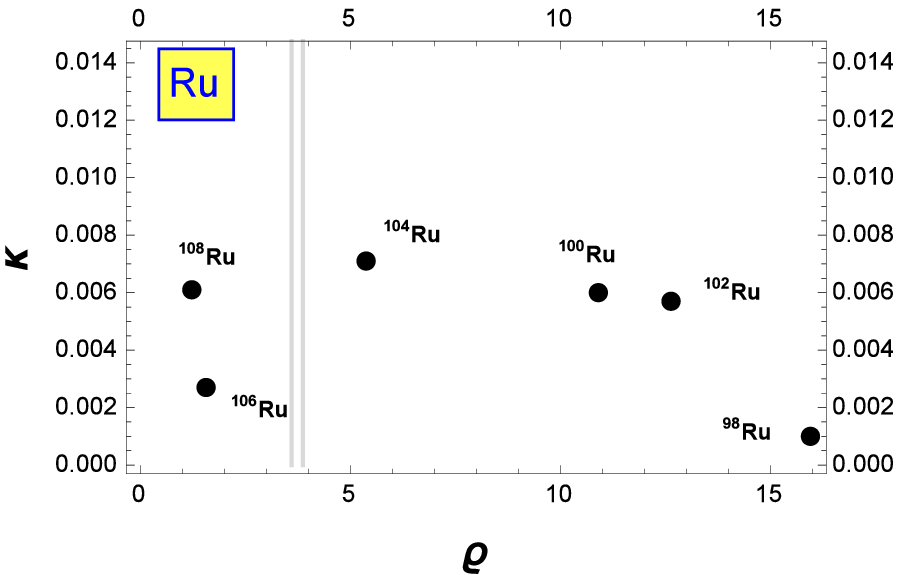}}{}
	\makeatother 
	\caption{{The ML parameter $ \kappa $ as a function of $ \varrho $ for isotopes of Ru, Xe, Pt and Nd. Gray vertical lines indicate the critical point region .}}
	\label{Isotropic2}
\end{figure}
\egroup

The variations of the ML parameter $ \kappa $ as a function of $ \varrho $ are shown in \fref{Isotropic2} for the most isotopic chains considered in this work. By analyzing this figure, one can see that, excepting the nuclei $^{128-130}$Xe and $^{192}$Pt with the ML parameter equal to 0, the ML parameter $ \kappa $ is spread between 0.0010 and 0.0071, between 0.0025 and 0.0074, between 0.0020 and 0.0132 and between 0.0002 and 0.0105 for Ru, Nd, Xe and Pt isotopes, respectively. Thus, unlike Pt isotopes, whose smallest value of $ \kappa $ corresponds to $^{194}$Pt, for the other isotopic chains, the smallest value of $ \kappa $ corresponds to the lightest nuclei of each isotopes.

On the other hand, the results presented in \fref{Isotropic2} are in concordance with the \fref{Isotropic} where the shape of the studied nuclei is easily determined. Indeed, the regions above (spherical shape) and bellow (deformed shape) the dashed lines  in \fref{Isotropic} are equivalent to the right  and  left sides, respectively, of the critical region between the gray vertical lines In \fref{Isotropic2} .
\subsection{\label{sec:transition} B(E2) transition rates}
Before the treatment of B(E2) transition probabilities, it is important to analyze the corrected wave functions  as well as the corresponding density probability distribution.
The behaviors of these two quantities versus $ y $ for values of the sextic potential parameters  $\varrho = 2 $, 6 and for $\kappa$ values varying from 0 to 0.3 are depicted in figures \ref{figfct} and \ref{figdens} for g.s. band labeled by   $ \eta^{Corr}_{0,0} $ and for the first excited $ \beta $ state labeled by $ \eta^{Corr}_{1,0} $, respectively.
\bgroup
\begin{figure}
	\begin{center}
		{\includegraphics[scale=0.80]{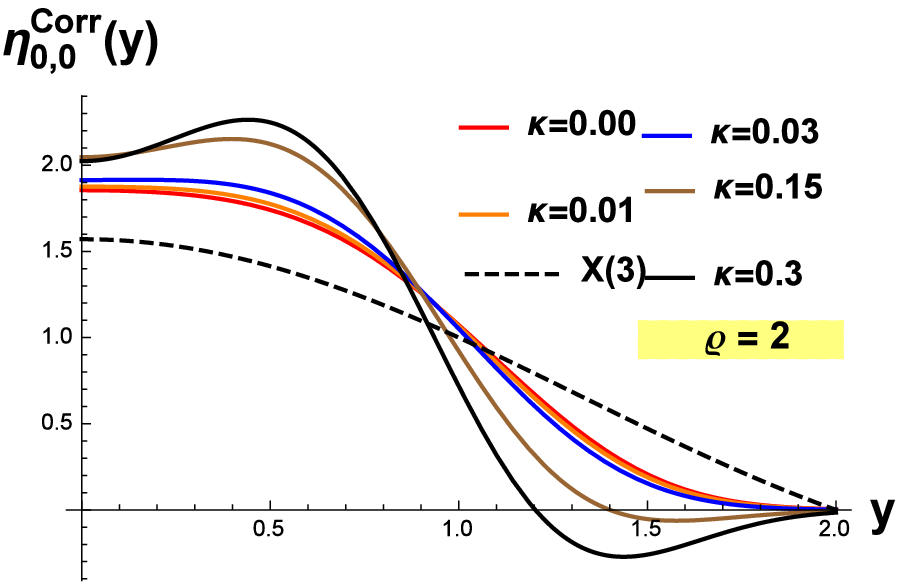}}{}{\includegraphics[scale=0.80]{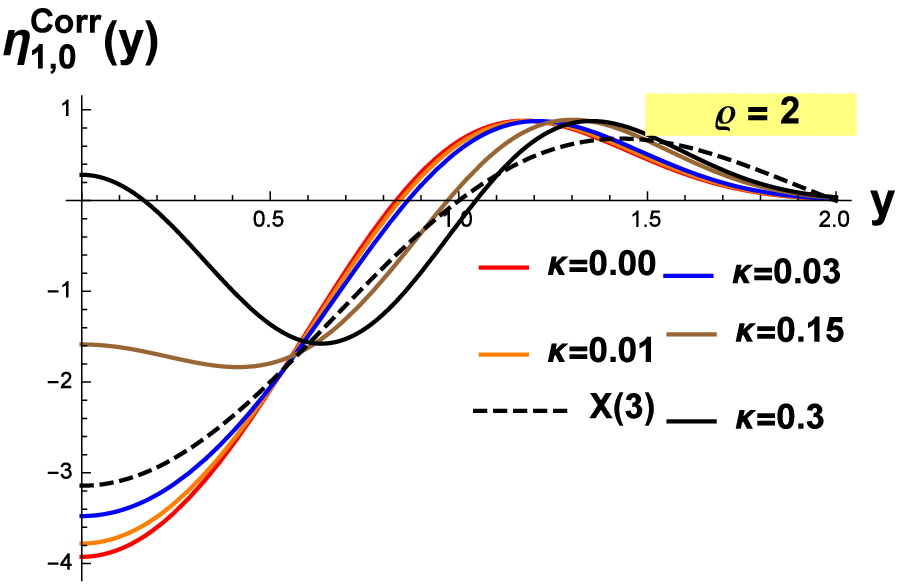}}{}\\
		{\includegraphics[scale=0.80]{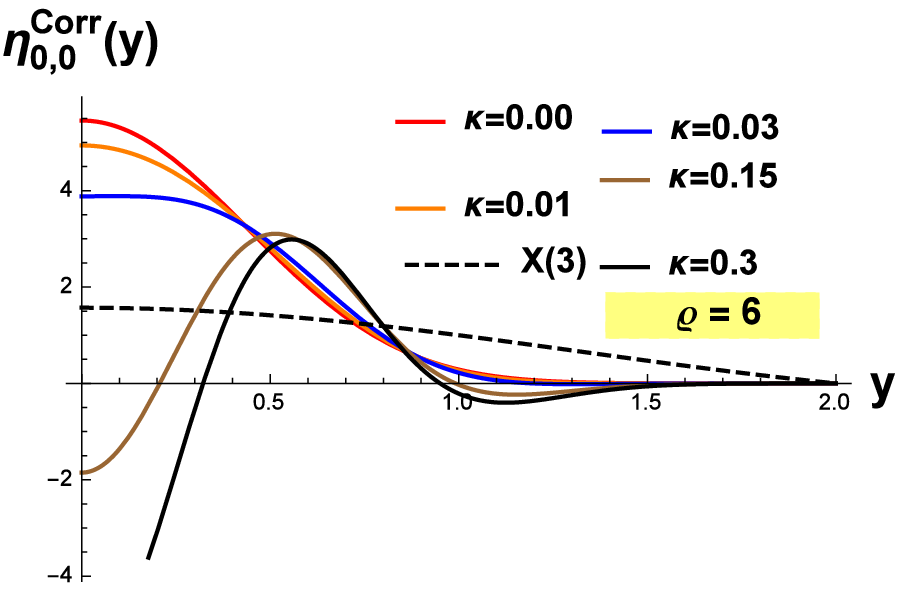}}{}{\includegraphics[scale=0.80]{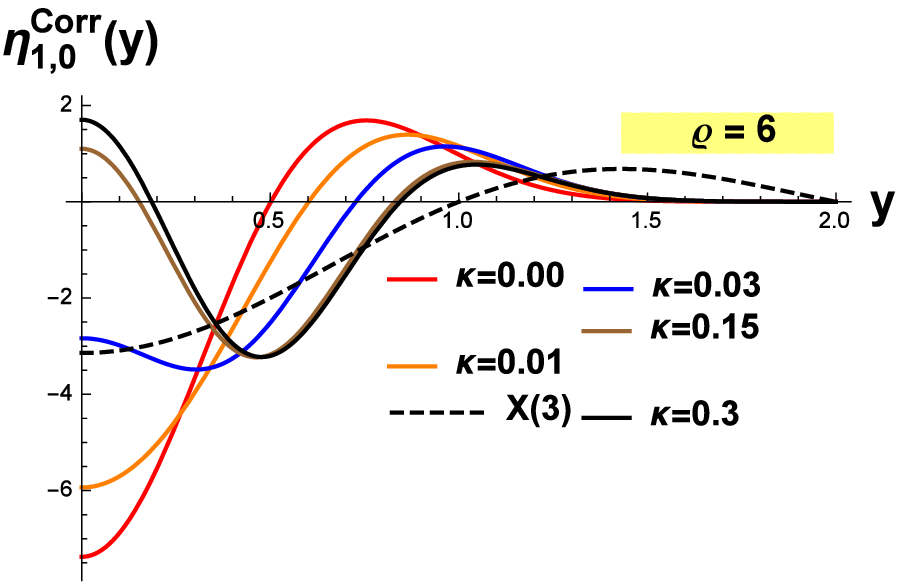}}{}
		\caption{{  The corrected wave function drawn as a function of  $ y $  and free parameters $ \varrho $  and $ \kappa $  for ground state labeled by   $ \eta^{Corr}_{0,0} $ and for the first excited $ \beta $ state labeled by $ \eta^{Corr}_{1,0} $  for the case  $ \varrho = 2 $ (Upper panel) and for  $ \varrho = 6 $ (Lower panel) }}
		\label{figfct}
	\end{center}
\end{figure}
\egroup

\bgroup
\begin{figure}
	\begin{center}
		{\includegraphics[scale=0.80]{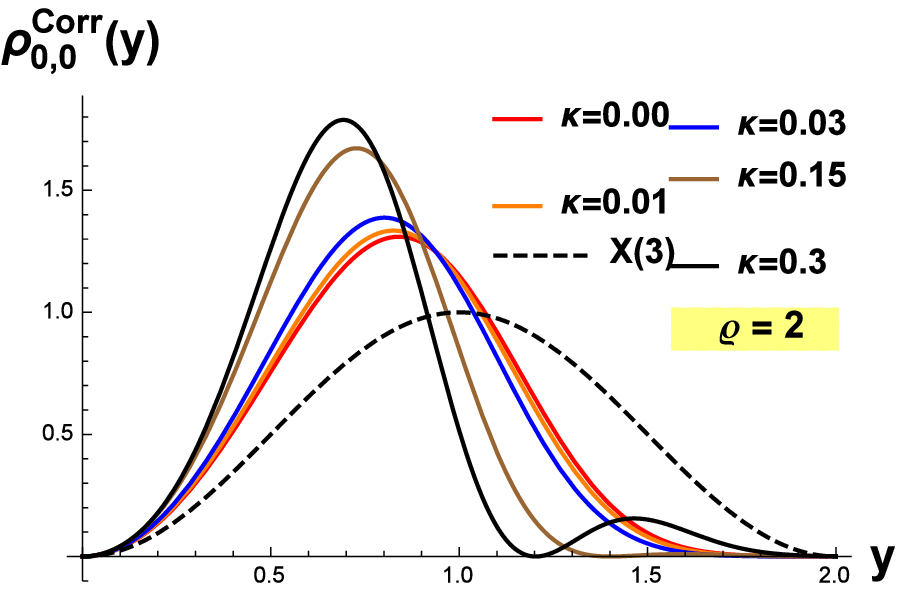}}{}{\includegraphics[scale=0.80]{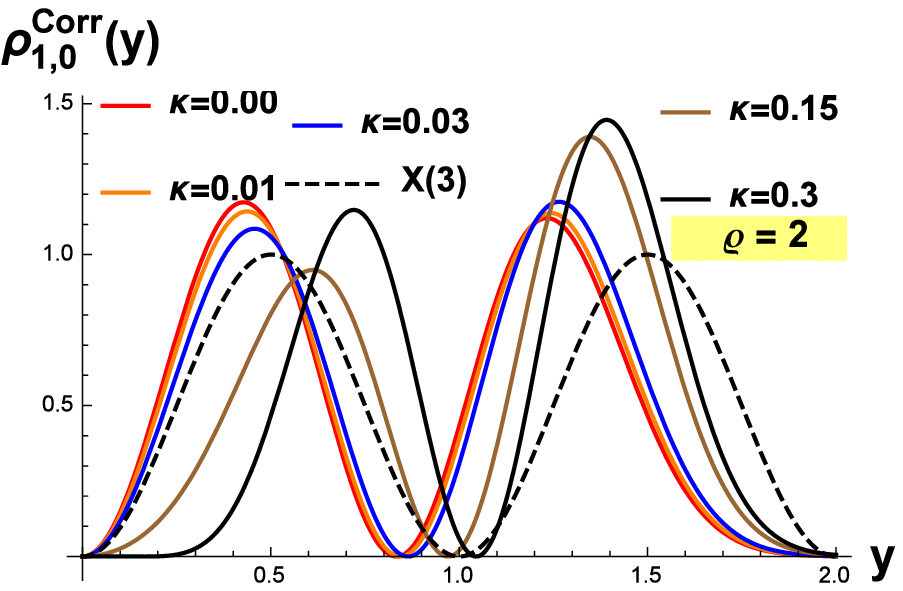}}{}\\
		{\includegraphics[scale=0.80]{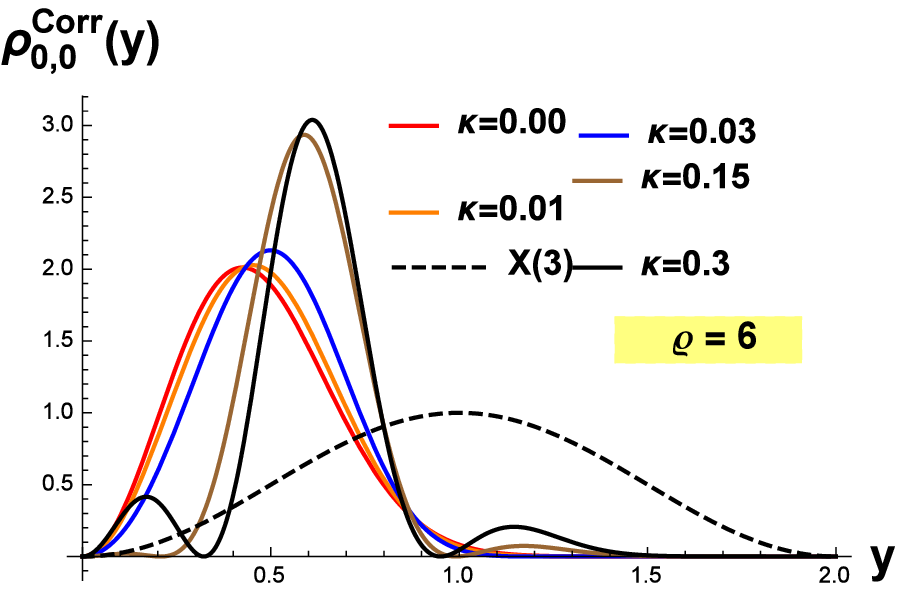}}{}{\includegraphics[scale=0.80]{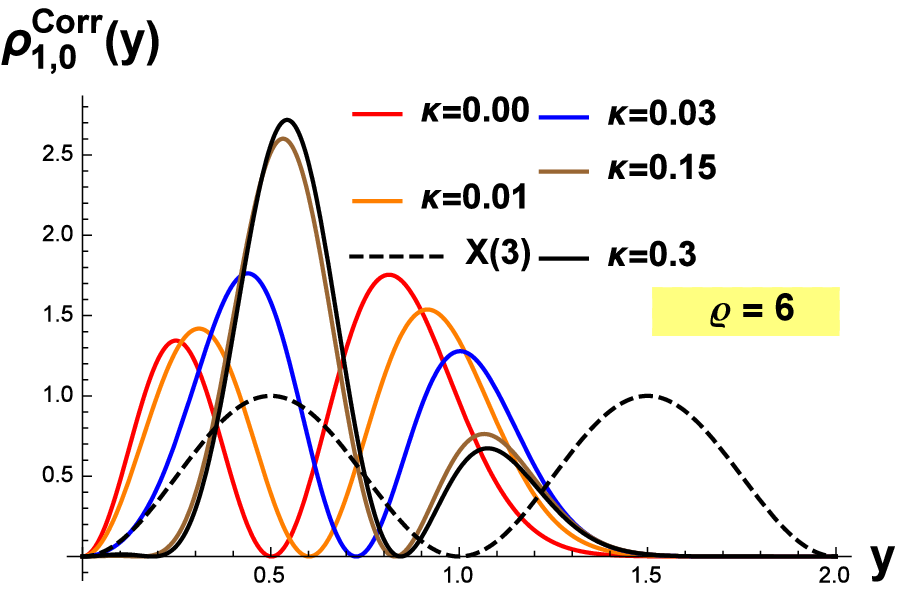}}{}	
		\makeatother 
		\caption{{The density of probability distribution  as a function of  $ y $  and free parameters $ \varrho $ and $ \kappa $  for ground state labeled by   $ \rho^{Corr}_{0,0} $ and for the first excited $ \beta $ state labeled by $ \rho^{Corr}_{1,0} $ for the case $ \varrho = 2 $    (Upper panel)  and for $ \varrho = 6 $  (Lower panel).}}
		\label{figdens}
	\end{center}
\end{figure}
\egroup
 From \fref{figfct}, one can see that for values of  $ \kappa $ up to  0.03, the behavior of the wave function is regular corresponding well and truly to the ground state.
Nevertheless, beyond this limit value, the wave function starts showing a local extremum which means that we are getting away from the ground state. The same phenomenon occurs in the excited $ \eta^{Corr}_{1,0} $ state. As a result the values ($\kappa  =0.03 $) shows the limitation of the perturbation term.
Moreover, from \fref{figdens}, one can see that when $\kappa$ is raised, in the case of  $\varrho$ = 2  (deformed shape), the density probability peaks are shifted backward on the left side in respect to the X(3) peak, while in the case of $\varrho$ = 6   (spherical shape), they are shifted forward on the right side. Also the full width at half maximum of the density probability distribution is the smallest in the presence of a ML compared to its absence.

The B(E2) transition rates are given in Tables \ref{tabpro1}, \ref{tabpro3} and \ref{tabpro2}. The agreement of\\ X(3)-SML with the corresponding experimental data is good taking into account the fact that the two free parameters $ \varrho $  and $ \kappa $ were  fitted only for the energy spectra. We notice that there are some differences in the g.s. band of some nuclei for which X(3)-SML predicts increasing values in respect to L while experimental data manifest a decreasing trend. This problem can be partly removed by including anharmonicities in the transition operator \eref{opper} as in \cite{b37}.
We can also see that the ML improve the B(E2) results of the interband transitions  from $ \beta_{1} $ band to g.s. band.
\newpage
\begin{table}[ht!]
	\caption{ Several B(E2) transition rates normalized as in $\eref{eq39} $ are compared with the available experimental data \cite{b38,b39,b40,b41,b42,b43,b52,b53,b56}  and with results available in \cite{b29} of the nuclei $^{98-108}$Ru, $^{100-102}$Mo, $^{132-134}$Ce and $^{150}$Sm. In the first line of each nucleus are given the experimental data, in the second line are  the corresponding X(3)-S \cite{b29}, while in third line are the corresponding theoretical results. }	
	\begin{center}
		\footnotesize	
		\begin{tabular}{ c c c c c c c c c c c}
			\br  
			Nucleus\qquad& $ \frac{4_{1}\rightarrow2_{1}}{2_{1}\rightarrow0_{1}} $ & $\frac{6_{1}\rightarrow4_{1}}{2_{1}\rightarrow0_{1}} $ & $\frac{8_{1}\rightarrow6_{1}}{2_{1}\rightarrow0_{1}} $ & $\frac{10_{1}\rightarrow8_{1}}{2_{1}\rightarrow0_{1}} $ &  $\frac{0_{\beta}\rightarrow2_{1}}{2_{1}\rightarrow0_{1}} $ & $\frac{2_{\beta}\rightarrow2_{1}}{2_{1}\rightarrow0_{1}}$& $\frac{2_{\beta}\rightarrow4_{1}}{2_{1}\rightarrow0_{1}} $ & 
			$\frac{2_{\beta}\rightarrow0_{\beta}}{2_{1}\rightarrow0_{1}} $ && \\
			\mr
			{$^{98}$Ru} \quad{Exp}&0.38(11)& 0.40(8)& 0.08(2)&&&&&&&\\
			\quad\qquad{X(3)-S}&2.34& 3.48& 4.62& 5.62& 3.21& 0.21& 1.47& 1.52&&\\
			\quad\qquad{X(3)-SML}&2.44&3.86& 5.28& 5.94& 2.92& 0.09&  1.39& 1.56&& \\
			\mr
			{$^{100}$Ru} \quad{Exp}&1.43(11)& 4.78(5)&&& 0.98(14)&&&&&\\
			\quad\qquad{X(3)-S}&2.23& 3.08& 3.98& 4.63& 2.83& 0.18& 1.02& 1.24&&\\
			\quad\qquad{X(3)-SML }&2.57& 4.45& 5.44& 5.63& 1.69& 0.03& 0.76& 1.34&& \\
			\mr
			{$^{102}$Ru} \quad{ Exp}&1.48(25)& 1.52(56)& 1.26(43)& 1.28(47)& 0.78(14)&&&&&\\
			\quad\qquad{X(3)-S}&2.25& 3.15& 4.08& 4.77& 2.90& 0.19& 1.09& 1.28&&\\
			\quad\qquad{X(3)-SML}&2.59& 4.50& 5.41& 5.69& 1.56& 0.05& 0.77& 1.33&& \\
			\mr				
			{$^{104}$Ru} \quad{ Exp}&1.43(16)&&& 0.43(5)&&&&&&\\
			\quad\qquad{X(3)-S}&2.18& 2.94& 3.76& 4.31& 2.65& 0.17& 0.88& 1.15&&\\
			\quad\qquad{X(3)-SML}&2.42& 3.71& 4.79&4.97& 2.16& 0.01& 0.73& 1.22 && \\
			\mr
			{$^{106}$Ru} \quad{Exp}&&&&&&&&&&\\
			\quad\qquad{X(3)-S}&1.75& 2.07& 2.47& 2.67& 1.22& 0.09& 0.24& 0.69&&\\
			\quad\qquad{X(3)-SML}&1.86& 2.27& 2.76& 3.00& 1.56& 0.10& 0.34& 0.77&& \\
			\mr				
			{$^{108}$Ru} \quad{Exp}&1.65(20)&&&&&&&&&\\
			\quad\qquad{X(3)-S}&1.66& 1.93& 2.27& 2.43& 0.95& 0.08& 0.16& 0.64&&\\
			\quad\qquad{X(3)-SML}&1.79&2.16&2.61& 2.81&1.39&0.09& 0.28& 0.72&&\\
			\mr
			{$^{100}$Mo} \quad{Exp}&1.86(11)& 2.54(38)& 3.32(49)&& 2.49(12)&$ \approx 0 $& 0.97(49)& 0.38(11)&&\\
			\quad\qquad{X(3)-S}&2.35& 3.49& 4.64& 5.64& 3.21& 0.21& 1.48& 1.53&&\\
			\quad\qquad{X(3)-SML}&2.46& 4.00& 5.49& 6.06& 2.83& 0.06& 1.38& 1.58&& \\
			\mr		
			{$^{102}$Mo} \quad{Exp}&1.20(28)&&& 0.95(42)&&&&&&\\
			\quad\qquad{X(3)-S}&2.05& 2.64& 3.31& 3.71& 2.23& 0.15& 0.62& 0.97&&\\
			\quad\qquad{X(3)-SML}&2.46& 3.86& 4.80& 4.93& 1.78& 0.004& 0.57& 1.17&& \\
			\mr
			{$^{132}$Ce} \quad{Exp}&0.75(17)& 0.27(10)& 0.47(4)& 0.07(2)&&&&&&\\
			\quad\qquad{X(3)-S}&2.00& 2.53& 3.15& 3.51& 2.06& 0.14& 0.54& 0.91&&\\
			\quad\qquad{X(3)-SML}&2.63& 5.98& 3.47& 6.54& 0.23& 0.69&1.01& 1.60&&\\
			\mr	
			{$^{134}$Ce} \quad{Exp}&0.75(17)& 0.27(10)& 0.47(4)& 0.07(2)&&&&&&\\
			\quad\qquad{X(3)-S}&1.91& 2.35& 2.88& 3.17& 1.75& 0.12& 0.41& 0.82&&\\
			\quad\qquad{X(3)-SML}&1.95& 2.45& 3.03& 3.34& 1.88& 0.12& 0.46& 0.86&&\\
			\mr
			{$^{150}$Sm} \quad{Exp}&1.93(30)& 2.63(88)& 2.98(158)& & 0.93(9)& & & $1.93^{+0.70}_{-0.53}$&&\\
			\quad\qquad{X(3)-S}&2.29& 3.28& 4.30& 5.09& 3.03& 0.20& 1.23&1.37&&\\
			\quad\qquad{X(3)-SML}&2.55& 4.36& 5.73& 5.90& 2.15& $ \approx 0 $& 1.01& 1.46&&\\
			\br	
		\end{tabular}
	\end{center}
	\label{tabpro1}
\end{table}
\normalsize
\newpage
\begin{table}[ht!]
	\caption{The same as in \Tref{tabpro1}, but for the available experimental data\cite{b44,b45,b46,b47,b48,b54,b55,b56,b57,b58,b59,b60} of the nuclei $^{152}$Sm, $^{146-150}$Nd, $ ^{154}$Gd, $ ^{156}$Dy, $^{172}$Os and $^{116-124}$Xe.}
	\begin{center}
		\footnotesize	
		\begin{tabular}{ c c c c c c c c c c c}
			\br  
			Nucleus\qquad& $ \frac{4_{1}\rightarrow2_{1}}{2_{1}\rightarrow0_{1}} $ & $\frac{6_{1}\rightarrow4_{1}}{2_{1}\rightarrow0_{1}} $ & $\frac{8_{1}\rightarrow6_{1}}{2_{1}\rightarrow0_{1}} $ & $\frac{10_{1}\rightarrow8_{1}}{2_{1}\rightarrow0_{1}} $ &  $\frac{0_{\beta}\rightarrow2_{1}}{2_{1}\rightarrow0_{1}} $ & $\frac{2_{\beta}\rightarrow2_{1}}{2_{1}\rightarrow0_{1}}$& $\frac{2_{\beta}\rightarrow4_{1}}{2_{1}\rightarrow0_{1}} $ & 
			$\frac{2_{\beta}\rightarrow0_{\beta}}{2_{1}\rightarrow0_{1}} $ && \\	
			\mr
			{$^{152}$Sm} \quad{Exp}&1.44(2)& 1.66(3)& 2.02(4)& $2.17^{+0.24}_{-0.18}$& 0.23(1)& 0.04& 0.12(1)& 1.17(8)&&\\
			\quad\qquad{X(3)-S}&1.75& 2.07& 2.47& 2.67& 1.22& 0.09& 0.24& 0.69&&\\
			\quad\qquad{X(3)-SML}&2.33& 3.43& 4.25& 4.37& 1.71& $ \approx0 $& 0.44& 1.03&&\\	
			\mr
			{$^{146}$Nd} \quad{Exp}&$1.47(39)$&&&&&&&&&\\
			\quad\qquad{X(3)-S}&2.30& 3.31& 4.35& 5.17& 3.06& 0.20& 1.27& 1.40&&\\
			\quad\qquad{X(3)-SML}&2.50& 4.11& 5.52& 5.79& 2.43& 0.010& 1.09& 1.47&&\\
			\mr
			{$^{148}$Nd} \quad{Exp}&1.62(9)& 1.76(14)& 1.69(30)&& 0.54(4)& 0.25(3)& 0.28(14)&&&\\
			\quad\qquad{X(3)-S}&2.22& 3.06& 3.94& 4.57& 2.80& 0.18& 1.00& 1.23&&\\
			\quad\qquad{X(3)-SML}&2.31& 3.31& 4.34& 4.83& 2.68& 0.11& 0.97& 1.27&&\\
			\mr
			{$^{150}$Nd} \quad{Exp}&1.56(4)& 1.78(9)& 1.86(20)& 1.73(10)& 0.37(2)& 0.09(3)& 0.16(6)& 1.38(112)&&\\
			\quad\qquad{X(3)-S}&1.66& 1.93& 2.27& 2.43& 0.95& 0.08& 0.16& 0.64&&\\
			\quad\qquad{X(3)-SML}&1.87& 2.30& 2.81& 3.01&1.56& 0.09& 0.33& 0.76&&\\	
			\mr
			{$^{154}$Gd} \quad{Exp}&1.56(6)& 1.82(10)& 1.99(11)& 2.29(26)& 0.33(5)& 0.04& 0.12(1)& 0.62(6)&&\\
			\quad\qquad{X(3)-S}&1.71& 2.01& 2.39& 2.57& 1.11& 0.09& 0.20& 0.67&&\\
			\quad\qquad{X(3)-SML}&2.35& 3.49& 4.25& 4.37& 1.56& 0.004& 0.38& 1.01&&\\	
			\mr
			{$^{156}$Dy} \quad{Exp}&1.63(2)& 1.87(6)& 2.07(20)& 2.20(27)&& 0.09(3)& 0.08(3)&&&\\
			\quad\qquad{X(3)-S}&1.71& 2.02& 2.40& 2.58& 1.12& 0.09& 0.21& 0.67&&\\
			\quad\qquad{X(3)-SML}&2.33& 3.43& 4.24& 4.35& 1.67&  $ \approx0 $& 0.42& 1.02&&\\
			\mr
			{$^{172}$Os} \quad{Exp}&1.50(17)& 2.61(38)& 3.30(98)& 1.65(36)&&&&&&\\
			\quad\qquad{X(3)-S}&2.21& 3.03& 3.90& 4.51& 2.76& 0.18& 0.97& 1.21&&\\
			\quad\qquad{X(3)-SML}&2.59& 4.59& 5.32& 5.62& 1.41& 0.08& 0.69& 1.32&&\\
			\mr
			{$^{116}$Xe} \quad{Exp}&1.75(11)& 1.58(15)&& 1.56(29)&&&&&&\\
			\quad\qquad{X(3)-S}&2.32& 3.38& 4.45& 5.34& 3.12& 0.20& 1.35& 1.44&&\\
			\quad\qquad{X(3)-SML}& 2.49& 4.08& 5.55& 5.92& 2.59& 0.03& 1.21& 1.52&& \\	
			\mr
			{$^{118}$Xe} \quad{Exp}&1.11(6)& 0.88(23)& 0.49(18)& $ > 0.7 $&&&&&&\\
			\quad\qquad{X(3)-S}&2.22& 3.06& 3.95& 4.58& 2.80& 0.18& 1.00& 1.23&&\\
			\quad\qquad{X(3)-SML}&2.62& 4.91& 5.13& 5.75& 1.05& 0.18& 0.68&1.35&& \\
			 \mr	
			 {$^{120}$Xe} \quad{Exp}&1.16(10)& 1.17(19)& 0.96(17)& 0.91(16)&&&&&&\\
			 \quad\qquad{X(3)-S}&2.19& 2.98& 3.82& 4.40& 2.70& 0.18& 0.91& 1.17&&\\
			 \quad\qquad{X(3)-SML}&2.62& 5.19& 4.57& 5.82& 0.64& 0.33& 0.68& 1.37&& \\
			 \mr	
			 {$^{122}$Xe} \quad{Exp}&1.46(11)& 1.41(9)& 1.03(8)& 1.54(10)&&&&&&\\
			 \quad\qquad{X(3)-S}&1.95& 2.44& 3.01& 3.33& 1.90& 0.13& 0.47& 0.86&&\\
			 \quad\qquad{X(3)-SML}&2.08& 2.71& 3.41& 3.77& 2.20& 0.12& 0.61& 0.98&&\\
			 \mr 
			 {$^{124}$Xe} \quad{Exp}&1.17(4)& 1.52(14)& 1.14(36)& 0.36(5)&&&&&&\\
			 \quad\qquad{X(3)-S}&2.10& 2.76& 3.48& 3.94& 2.40& 0.16& 0.72& 1.04&&\\
			 \quad\qquad{X(3)-SML}&2.62& 5.29& 4.18& 5.86& 0.48& 0.42& 0.70&1.38&&\\ 
			\br	
		\end{tabular}
	\end{center}
	\label{tabpro2}
\end{table}
\normalsize
\newpage
\begin{table} 	
	\caption{The same as in \Tref{tabpro1}, but for the available experimental data \cite{b49,b51,b61,b62,b63,b64,b65,b66,b68,b69} of the nuclei $^{126,130}$Xe and $^{180-196}$Pt.}
	
	\begin{center}
		\footnotesize
		\begin{tabular}{ c c c c c c c c c c c}  
			\br
			Nucleus\qquad& $ \frac{4_{1}\rightarrow2_{1}}{2_{1}\rightarrow0_{1}} $ & $\frac{6_{1}\rightarrow4_{1}}{2_{1}\rightarrow0_{1}} $ & $\frac{8_{1}\rightarrow6_{1}}{2_{1}\rightarrow0_{1}} $ & $\frac{10_{1}\rightarrow8_{1}}{2_{1}\rightarrow0_{1}} $ &  $\frac{0_{\beta}\rightarrow2_{1}}{2_{1}\rightarrow0_{1}} $ & $\frac{2_{\beta}\rightarrow2_{1}}{2_{1}\rightarrow0_{1}}$& $\frac{2_{\beta}\rightarrow4_{1}}{2_{1}\rightarrow0_{1}} $ & 
			$\frac{2_{\beta}\rightarrow0_{\beta}}{2_{1}\rightarrow0_{1}} $ && \\
		   
			\mr	
			{$^{126}$Xe} \quad{Exp}&&&&&&&&&&\\
			\quad\qquad{X(3)-S}&2.20& 2.99& 3.83& 4.42& 2.71& 0.18& 0.93& 1.18&&\\
			\quad\qquad{X(3)-SML}&2.63& 5.12& 4.71& 5.80& 0.73& 0.29& 0.67& 1.36&&\\
			\mr
			{$^{128}$Xe} \quad{ Exp}&1.47(15)& 1.94(20)& 2.39(30)&&&&&&&\\
			\quad\qquad{X(3)-S}&2.21& 3.03& 3.89& 4.50& 2.76& 0.18& 0.97& 1.20&&\\
			\quad\qquad{X(3)-SML}&2.21& 3.03& 3.89& 4.50& 2.76& 0.18& 0.97& 1.20&&\\
			\mr
			{$^{130}$Xe} \quad{Exp}&&&&&&&&&&\\
			\quad\qquad{X(3)-S}&2.28& 3.23& 4.22& 4.97& 2.99& 0.19& 1.18& 1.34&&\\
			\quad\qquad{X(3)-SML}&2.28& 3.23& 4.22& 4.97& 2.99& 0.19& 1.18& 1.34&&\\
			\mr
			{$^{180}$Pt} \quad{ Exp}&0.92(22)& $\geq 0.29$&&&&&&&&\\
			\quad\qquad{X(3)-S}&1.78& 2.12& 2.55& 2.77& 1.33& 0.10& 0.27& 0.72&&\\
			\quad\qquad{X(3)-SML}&2.19& 3.00& 3.80& 4.02& 2.04& 0.04& 0.54& 1.00&& \\
			\mr
			{$^{182}$Pt} \quad{ Exp}&&&&&&&&&&\\
			\quad\qquad{X(3)-S}&1.79& 2.14& 2.57& 2.80& 1.36& 0.10& 0.28& 0.72&&\\
			\quad\qquad{X(3)-SML}&2.29& 3.29& 4.18& 4.35& 2.00& 0.02& 0.56& 1.07&& \\
			\mr
			{$^{184}$Pt} \quad{Exp}&1.65(9)& 1.78(12)& 2.13(16)& 2.44(33)&&&&&&\\
			\quad\qquad{X(3)-S}&1.81& 2.17& 2.62& 2.86& 1.42& 0.10& 0.30& 0.74&&\\
			\quad\qquad{X(3)-SML}&2.20& 3.03& 3.86& 4.09& 2.09& 0.05& 0.57& 1.03&& \\
			\mr
			{$^{186}$Pt} \quad{Exp}&&& &&&&&&&\\
			\quad\qquad{X(3)-S}&2.08& 2.70& 3.40& 3.83& 2.32& 0.15& 0.67& 1.01&&\\
			\quad\qquad{X(3)-SML}&2.61& 5.64& 3.82& 6.17& 0.33& 0.56& 0.85& 1.49&& \\
			\mr	
			{$^{188}$Pt} \quad{Exp}&&& &&&&&&&\\
			\quad\qquad{X(3)-S}&2.01& 2.55& 3.17& 3.54& 2.08& 0.14& 0.55& 0.92&&\\
			\quad\qquad{X(3)-SML}&2.39& 3.59& 4.60& 4.77& 2.09& 0.01& 0.66& 1.17&& \\
			\mr	
			{$^{190}$Pt} \quad{Exp}&&& &&&&&&&\\
			\quad\qquad{X(3)-S}&2.10& 2.75& 3.46& 3.92& 2.39& 0.16& 0.71& 1.03&&\\
			\quad\qquad{X(3)-SML}&2.41& 3.68& 4.75& 4.93& 2.14& 0.01& 0.71& 1.22&& \\
			\mr	
		{$^{192}$Pt} \quad{ Exp}&1.56(9)& 1.22(53)&&&&&&&&\\
			\quad\qquad{X(3)-S}&2.00& 2.53& 3.14& 3.50& 2.05& 0.14& 0.54& 0.91&&\\
			\quad\qquad{X(3)-SML}&2.00& 2.53& 3.14& 3.50& 2.05& 0.14& 0.54& 0.91&&\\
			\mr			
			{$^{194}$Pt} \quad{Exp}&1.73(11)& 1.36(43)& 1.02(29)& 0.69(18)&&&&&&\\
			\quad\qquad{X(3)-S}&1.98& 2.49& 3.09& 3.43& 1.99& 0.14& 0.51& 0.89&&\\
			\quad\qquad{X(3)-SML}&2.03& 2.60& 3.25& 3.64& 2.16& 0.14& 0.59& 0.95&& \\
			\mr	
			{$^{196}$Pt} \quad{Exp}&1.48(2)& 1.80(10)& 1.92(25)&&&& $ \approx $ 0& 0.12(12)&&\\
			\quad\qquad{X(3)-S}&2.14& 2.83& 3.60& 4.10& 2.51& 0.17& 0.79& 1.09&&\\
			\quad\qquad{X(3)-SML}&2.32& 3.34& 4.36& 4.71& 2.47& 0.07& 0.84& 1.21&& \\				
			\br
		\end{tabular}
	\end{center}
	\label{tabpro3}
\end{table}
\normalsize
\newpage
\subsection{\label{sec:Momentum inertia}The moment of inertia of the ground state}

In this last section, we discuss the ML effect on the variation of the moment of inertia. For this purpose, we analyze the plot of the moment of inertia $ \theta(L) $ as a function of angular momentum $L$.
Nevertheless, the calculation of this quantity is very difficult owing to its dependence on the angular velocity $ \omega $.
The relation which is very often used in practice and easily allowing to calculate the moment of inertia of the ground state as a function of the total angular momentum has the form \cite{b78}:
\begin{equation}
\theta(L)=\frac{R}{\omega}=\frac{1}{2} \frac{d E}{d R^{2}} \approx \frac{2 L-1}{E(L)-E(L-2)}, R^{2}=L(L+1).
\label{angularmomentum}
\end{equation}

\begin{figure}
	\begin{center}
		{\includegraphics[scale=0.60]{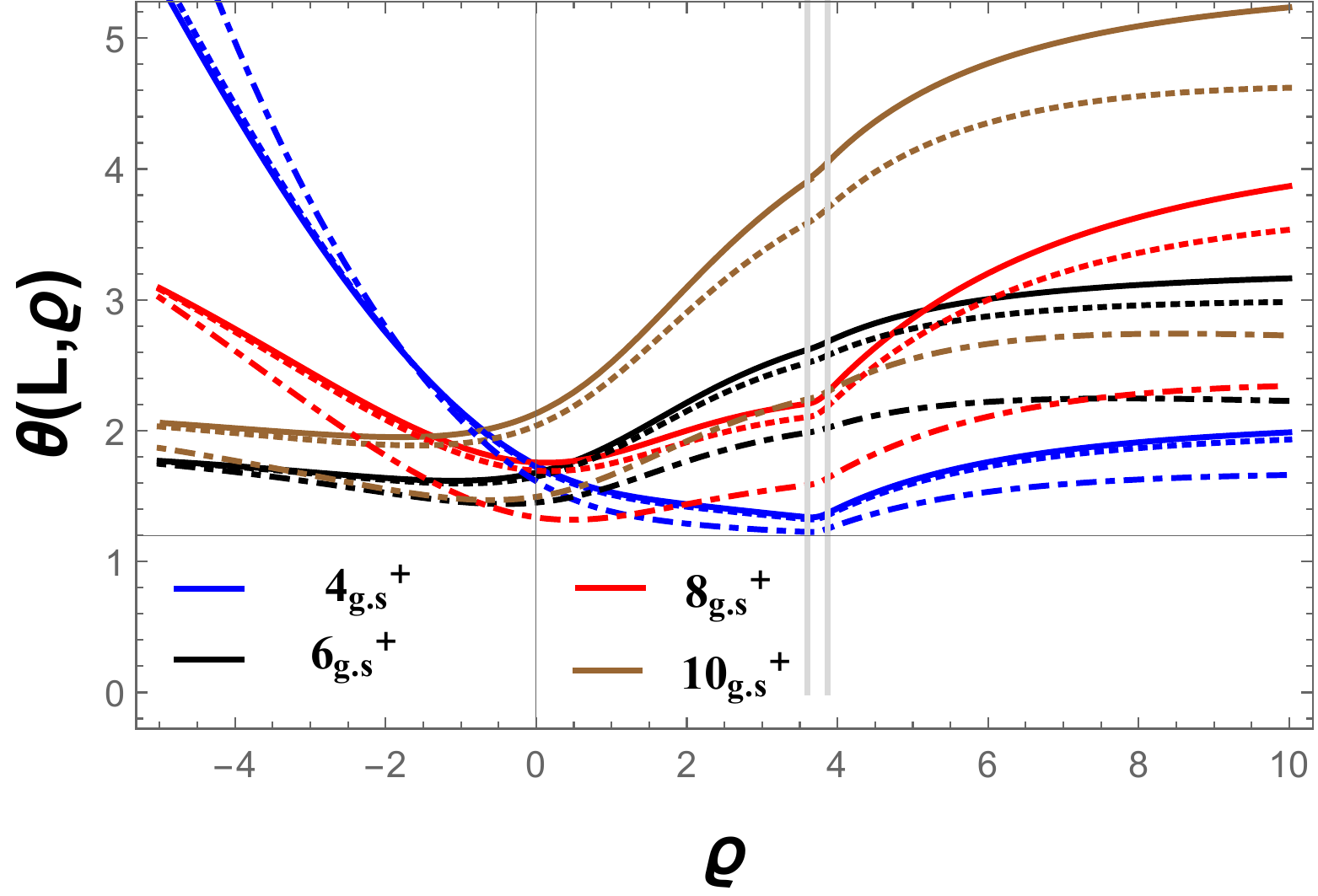}}{}\\
		\caption{{ Moment of inertia \eref{angularmomentum} for the g.s band normalized to the moment of the $ 2^{+} $ state for ML plotted as function of parameter $ \varrho $ for different values of the ML parameter: $ \kappa =0 $ (continuous curves), $ \kappa=0.001 $ (tiny-dashed curves),  and $ \kappa=0.01 $  (large-dashed curves).
				The blue, the black, the red and the brown curves correspond to $ 4^{+}_{g.s} $, $ 6^{+}_{g.s} $, $ 8^{+}_{g.s} $ and $ 10^{+}_{g.s} $, respectively.}}	
		\label{mom}
	\end{center}
\end{figure}

This relation is usually used to explain what is the so-called "Backbending" effect which corresponds to an apparent anomaly in the evolution of the angular momentum as a function of the frequency $\omega$.
A plot of the moment of inertia for the g.s band normalized to the moment of the $ 2^{+} $ state given by \Eref{angularmomentum} as a function of $\varrho$ is presented in \fref{mom}.

\begin{figure}
	{\includegraphics[scale=0.60]{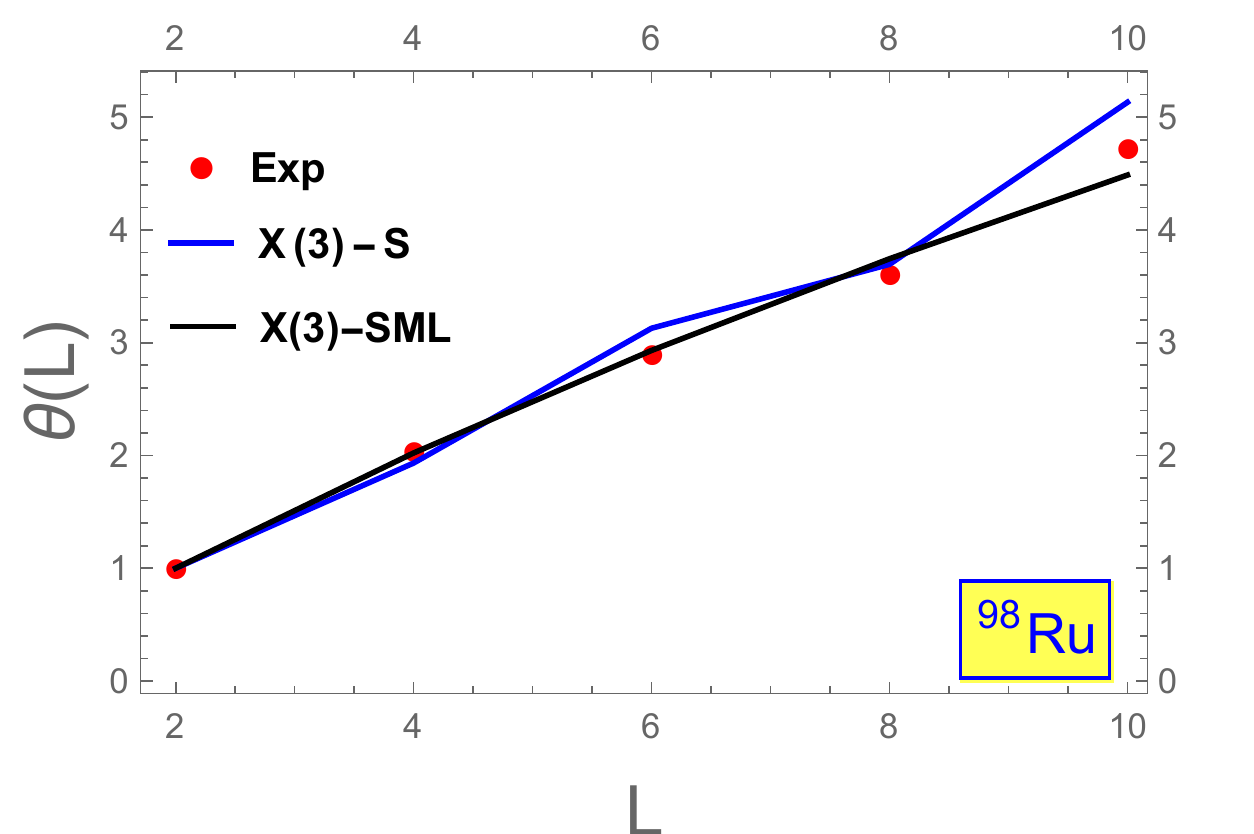}}{}{\includegraphics[scale=0.60]{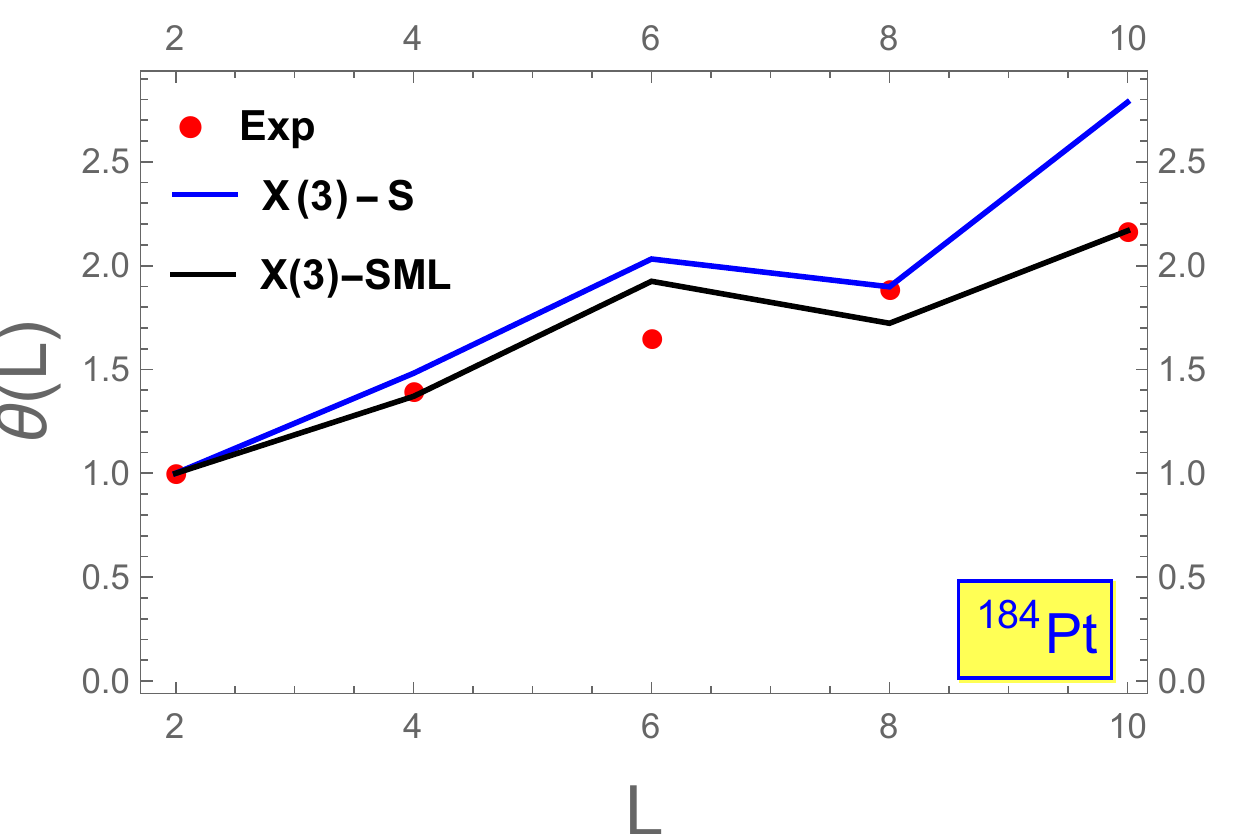}}{}\\	
	\makeatother 
	\caption{{Moment of inertia for the g.s band normalized to the moment of the $ 2^{+} $ state plotted as a function of L are compared with the available experimental data \cite{b38,b63} of the nuclei $^{98}$Ru (spherical region)  and $^{184}$Pt (deformed region). }}
	\label{MIexp}
\end{figure}
 For higher values of the parameter $\varrho$ defined in spherical region, $ \theta(L) $ is a monotonically increasing function of L, while for the other values of $\varrho$, especially for the deformed and the critical point regions, $ \theta(L) $ has not a monotonic behavior. This comes out from the fact that the moment of inertia depends on spacing between two successive levels. These two cases are clearly illustrated in \fref{MIexp}  for the nuclei $^{98}$Ru (spherical region) and $^{184}$Pt (deformed region).
Concerning the ML effect on the variation of the moment of inertia, from figure \eref{MIexp}, one can see that in the absence of ML ($\kappa  =0 $), the moment of inertia increases with L, while in its presence ($\kappa \ne 0 $), it takes moderate values removing a main drawback of the model and leading
to an improved agreement with the corresponding experimental data.

\section{Conclusion}
Here, we summarize the main results obtained in this work. We have solved the Bohr-Mottelson Hamiltonian in the $ \gamma$-rigid regime within the  Minimal Length (ML) formalism with sextic potential. The model is conventionally called $\mathrm{X}(3)$-SML. However, the Hamiltonian of the system is not soluble analytically for a potential other than the square well. In order to overcome such a difficulty, in the present work we have used a quasi-exact solvability (QES) and a quantum perturbation method (QPM). Finally, the energy spectrum and the wave functions are given in analytical forms which depend on two free parameters $\varrho$ and $\kappa$. The new elaborated model has allowed us to reproduce well the experimental data for energy ratios, transition rates and moments of inertia for several nuclei in comparison with the pure sextic model and others. Moreover, the additional effect of the ML on energy levels consisted of removing partially the degeneracy of states with different angular momenta \textsc{$\Delta L=2$}  belonging to different bands. The analysis of the corrected wave function, as well
as the probability density distribution, showed that the ML parameter has a physical
upper bound limit. Besides, it was shown that the ML has an important effect on the variation of the moment of inertia versus angular momentum inasmuch as it keeps such a variation moderate allowing to remove some drawback of the elaborated model in respect to others.

Before closing, we consider two perspectives of the present approach, namely: the first one is to consider the quasi-exact solvability orders $k>2$ for $\mathrm{X}(3)$-Sextic in the presence of ML and to verify whether the theoretical results will be improved more than in the case of k=2, the second one is to correct the energies and the wave functions to the second order instead of the first one, and to study its effect on the energy ratios and  transition rates.	
\bibliographystyle{elsarticle-num}

\end{document}